\documentclass[12pt]{article}
\usepackage{epsfig}
\usepackage{amssymb}

\def\a{\alpha}
\def\b{\beta}

\def\d{\delta}
\def\e{\epsilon}
\def\f{\phi}
\def\g{\gamma}

\def\j{\psi}
\def\k{\kappa}

\def\m{\mu}
\def\n{\nu}
\def\o{\omega}
\def\p{\pi}

\def\s{\sigma}

\def\D{\Delta}

\def\G{\Gamma}

\def\O{\Omega}

\def\ve{\varepsilon}

\def\dg{\dagger}                                     % hermitian conjugate
\def\mt{\widetilde{m}_1}
\def\mb{\overline{m}}
\def\th{\tilde{h}}
\def\beq{\begin{equation}}
\def\eeq{\end{equation}}

\def\bea{\begin{eqnarray}}
\def\eea{\end{eqnarray}}
\def\NO{\nonumber}

\def\pl#1#2#3{Phys.~Lett.~{\bf B {#1}} ({#2}) #3}
\def\np#1#2#3{Nucl.~Phys.~{\bf B {#1}} ({#2}) #3}
\def\prl#1#2#3{Phys.~Rev.~Lett.~{\bf #1} ({#2}) #3}
\def\pr#1#2#3{Phys.~Rev.~{\bf D {#1}} ({#2}) #3}

\begin{document}
\date{}
\title{
{\normalsize
\mbox{ }\hfill
\begin{minipage}{4cm}
DESY 03-100\\
UAB-FT-551 \\
CERN-TH/2003-199
\end{minipage}}\\
\vspace{1cm}
\bf Leptogenesis for Pedestrians}
\author{W.~Buchm\"uller \\
{\it Deutsches Elektronen-Synchrotron DESY, 22603 Hamburg, Germany}\\[2ex]
P.~Di Bari \\
{\it IFAE, Universitat Aut{\`o}noma de Barcelona,}\\
{\it 08193 Bellaterra (Barcelona), Spain}\\[2ex]
M.~Pl\"umacher\\
{\it Theory Division, CERN, 1211 Geneva 23, Switzerland}
}
\maketitle

\thispagestyle{empty}

%\centerline{\date{\today}}

\begin{abstract}

\noindent During the process of thermal leptogenesis temperature
decreases by about one order of magnitude while the baryon
asymmetry is generated. We present an analytical description of
this process so that the dependence on the neutrino mass
parameters becomes transparent. In the case of maximal $C\!P$
asymmetry all decay and scattering rates in the plasma are
determined by the mass $M_1$ of the decaying heavy Majorana
neutrino, the effective light neutrino mass $\mt$ and the absolute
mass scale $\mb$ of the light neutrinos. In the mass range
suggested by neutrino oscillations, $m_{\rm sol} \simeq 8\times
10^{-3}\ {\rm eV} \lesssim \mt \lesssim
 m_{\rm atm} \simeq 5\times 10^{-2}\ {\rm eV}$,
leptogenesis is dominated just by decays and inverse decays. The
effect of all other scattering processes lies within the
theoretical uncertainty of present calculations. The final baryon
asymmetry is dominantly produced at a temperature $T_B$ which can
be about one order of magnitude below the heavy neutrino mass
$M_1$. We also derive an analytical expression for the upper bound
on the light neutrino masses implied by successful leptogenesis.
\end{abstract}

\newpage

\section{Introduction and summary}

Leptogenesis \cite{fy86} provides a simple and elegant explanation of the cosmological
matter-antimatter asymmetry. A beautiful aspect of this mechanism is the connection
between the baryon asymmetry and neutrino properties. In its simplest version
leptogenesis is dominated by the $C\!P$ violating interactions of the lightest of the
heavy Majorana neutrinos, the seesaw partners of the ordinary neutrinos. The requirement
of successful baryogenesis yields stringent constraints on the masses of light and
heavy neutrinos. In particular, all light neutrino masses have to be smaller than
$0.1\,{\rm eV}$ \cite{bdp03}.

Leptogenesis is closely related to classical GUT baryogenesis \cite{kt}, where
the deviation of the distribution function of some heavy particle from its equilibrium
distribution provides the necessary departure from thermal equilibrium. The
non-equilibrium process of baryogenesis is usually studied by means of Boltzmann
equations \cite{kw80,hkx82}. In the same way, leptogenesis has been studied
during the past years, with increasing sophistication \cite{lut92}-\cite{gnx03}.
The goal of the present paper is to provide an analytical description of the
leptogenesis process such that the dependence on the neutrino mass parameters
becomes transparent. As we shall see this is important to understand the size of
corrections to the simplest Boltzmann equations, which have to
be taken into account to arrive eventually at a `theory of leptogenesis'.

We will first consider the  simplest case where the initial temperature $T_{\rm i}$ is larger
than $M_1$, the mass of the lightest heavy neutrino $N_1$. We will also
neglect decays of the two heavier neutrinos $N_2$ and $N_3$, assuming that
a generation of $B-L$ asymmetry from their decays either does not occur at all or
that it does not influence the final value of $B-L$. Further, we restrict ourselves to
the non supersymmetric case, and we assume that the lightest heavy neutrino $N_1$ is
the only
relevant degree of freedom beyond the standard model particle species.

Within this minimal framework the Boltzmann equations can be written in the
following form\footnote{We use the conventions of Ref.~\cite{bdp02}.},
\begin{eqnarray}
{dN_{N_1}\over dz} & = & -(D+S)\,(N_{N_1}-N_{N_1}^{\rm eq}) \;, \label{lg1} \\
{dN_{B-L}\over dz} & = & -\ve_1\,D\,(N_{N_1}-N_{N_1}^{\rm eq})-W\,N_{B-L} \;,\label{lg2}
\end{eqnarray}
where $z=M_1/T$. The number density $N_{N_1}$ and the amount of $B-L$ asymmetry, $N_{B-L}$,
are calculated in a portion of comoving volume that contains one photon at temperatures
$T\gg M_1$, so that the relativistic equilibrium $N_1$ number density is given by
$N_{N_1}^{\rm eq}(z\ll 1)=3/4$.
There are four classes of processes which contribute to the different terms in the
equations: decays, inverse decays, $\D L=1$ scatterings and $\D L=2$ processes mediated by
heavy neutrinos. The first three all modify the
$N_1$ abundance and try to push it towards its equilibrium value $N_{N_1}^{\rm eq}$.
Denoting by $H$ the Hubble expansion rate, the term $D = \Gamma_D/(H\,z)$
accounts for decays and inverse decays, while the scattering term $S = \Gamma_S/(H\,z)$
represents the $\D L=1$ scatterings. Decays also yield the source term for the generation
of the $B-L$ asymmetry, the first term in Eq.~(\ref{lg2}), while all other processes
contribute to the total washout term $W = \Gamma_W/(H\,z)$ which competes
with the decay source term. The expansion rate is given  by
\begin{equation}
H\simeq \sqrt{8\,\pi^3\,g_*\over 90} {M_1^2\over M_{\rm Pl}}\,{1\over z^{2}}
\simeq 1.66\,g_*\,{M_1^2\over M_{\rm Pl}}\,{1\over z^{2}}\;,
\end{equation}
where $g_*=g_{SM}=106.75$ is the total number of degrees of freedom, and
$M_{\rm Pl}=1.22\times10^{19}\,{\rm GeV}$ is the Planck mass. Note that we
have not included the $N_1$ degrees of freedom since, as we will see, in the preferred
strong washout regime, the heavy neutrinos are non-relativistic when the baryon
asymmetry is produced.

The two terms $D$ and $S$ depend on the {\em effective neutrino mass} \cite{plu97},
defined as
\begin{equation}\label{mtilde}
\widetilde{m}_1={(m_D^{\dagger}\,m_{D})_{11}\over M_1 }\; ,
\end{equation}
which has to be compared with the {\it equilibrium neutrino mass}
\begin{equation}\label{d}
m_* = {16\, \pi^{5/2}\,\sqrt{g_*} \over 3\,\sqrt{5}}\,
{v^2 \over M_{\rm Pl}} \simeq 1.08\times 10^{-3}\,{\rm eV}\;.
\end{equation}
The decay parameter
\begin{equation}\label{decpar}
K = {\G_D(z=\infty)\over H(z=1)} = {\widetilde{m}_1\over m_*} \;,
\end{equation}
introduced in the context of ordinary GUT baryogenesis \cite{kt},
controls whether or not $N_1$ decays are in equilibrium. Here
$\G_D(z=\infty)\equiv\widetilde{\G}_D$ is the $N_1$ decay width. The
washout term $W$ has two contributions, $W=W_0+\D W$; the first
term only depends on $\mt$, while the second one depends on the
product $M_1\,\mb^2$, where $\mb^2=m_1^2+m_2^2+m_3^2$ is the sum
of the light neutrino masses squared \cite{bdp02}.

The solution for $N_{B-L}$ is the sum of two terms \cite{kt},
\begin{equation}\label{solution}
N_{B-L}(z)=N_{B-L}^{\rm i}\,e^{-\int_{z_{\rm i}}^{z}\,dz'\,W(z')}
-{3\over 4}\,\ve_1\,\k(z;\mt,M_1\,\mb^2)\;,
\end{equation}
where the second term describes $B-L$ production from $N_1$ decays.
It is expressed in terms of the {\em efficiency factor} $\k$  \cite{bcx00} which does
not depend on the CP asymmetry $\ve_1$. In the following sections we shall use
two integral expressions for the efficiency factor,
\bea\label{ef}
\k(z)&=&{4\over 3} \int_{z_{\rm i}}^{z}\,dz'\,
D \left(N_{N_1}-N_{N_1}^{\rm eq}\right)\,e^{-\int_{z'}^{z}\,dz''\,W(z'')} \\
&=& - {4\over 3}\int_{z_{\rm i}}^{z}\,dz'\,
{D\over D+S}{dN_{N_1}\over dz'}\,e^{-\int_{z'}^{z}\,dz''\,W(z'')}\;.
\eea
Here $N_{N_1}$ and $dN_{N_1}/dz'$ are the solution of the first kinetic equation
(\ref{lg1}) and its derivative, respectively. The efficiency factor
$\k(z)$ is normalized in such a way
that its final value $\k_{\rm f} = \k(\infty)$ approaches one in the limit of thermal
initial abundance of the heavy neutrinos $N_1$ and no washout ($W=0$). In general, for
$N^{\rm i}_{N_1}\leq N_{N_1}^{\rm eq}=3/4$,
one has $\kappa_{\rm f}\leq 1$. The first term in Eq. (\ref{solution}) accounts
for the possible generation of a $B-L$ asymmetry before $N_1$ decays, e.g.
from decays of the two heavier neutrinos $N_2$ and $N_3$, or from a completely
independent mechanism. In the following we shall neglect such an initial asymmetry
$N_{B-L}^{\rm i}$. In \cite{bdp03} it was shown that for values $\mt > m_*$ even large
initial asymmetries are washed out
for initial temperatures $T_{\rm i}\gtrsim M_1$.

The predicted baryon to photon number ratio has to be compared with the value
$\eta_B$  measured at recombination. It is related to
$N_{B-L}^{\rm f} = N_{B-L}(z=\infty)$ by
$\eta_B=(a_{\rm sph}/f)\,N_{B-L}^{\rm f}$. Here $a_{\rm sph}=28/79$ \cite{ks88}
is the fraction of $B-L$
asymmetry converted into a baryon asymmetry by sphaleron processes,
and $f=N_{\g}^{\rm rec}/N_{\g}^{\star}=2387/86$ is the dilution factor
calculated assuming standard photon production from the onset
of leptogenesis till recombination. Using Eq. (\ref{ef}), one then obtains
\begin{equation}\label{etaB}
\eta_B = {3\over 4}\,{a_{\rm sph}\over f}\,\ve_1\,\k_{\rm f} \equiv d\,\ve_1\,\k_{\rm f}
\simeq 0.96\times 10^{-2}\,\ve_1\k_{\rm f}\;.
\end{equation}

In the following sections we will study analytically the solutions of the kinetic
equations, focusing in particular on the final value of the efficiency factor.
We start in sect.~2 with the basic framework of decays and inverse decays.
In the two regimes of weak ($\mt < m_*$) and strong ($\mt > m_*$) washout the
efficiency factor is obtained analytically, which then leads to a simple interpolation
valid for all values of $\mt$.
$\D L=1$ scatterings are added in sect.~3, and the resulting lower bounds on
the heavy neutrino mass $M_1$ and on the initial temperature $T_{\rm i}$ are
discussed. In sect.~4 an analytic derivation of the upper bound on the light
neutrino masses is given, and in sect.~5 various corrections are described which
have to be taken into account in a theory of leptogenesis.
In appendix A a detailed discussion of the $\D L=2$ processes in the resonance
region is presented. In the case of maximal $C\!P$ violation the
entire $\D L=2$ scattering cross section can be expressed in terms of $M_1$, $\mt$
and $\mb$. The resulting Boltzmann equations are compared with previously obtained
results based on exact Kadanoff-Baym equations. In appendix B various
useful formulae are collected.

Recently, two potentially important, and usually neglected,
effects on leptogenesis have been discussed: the $\D L=1$
processes involving gauge bosons \cite{pu03,gnx03} and thermal
corrections at high temperature \cite{gnx03}. Further, the strength of the
$\D L=2$ washout term has been corrected \cite{gnx03} compared to previous
analysis. However, the
reaction densities for the gauge boson processes are presently
controversial \cite{pu03,gnx03}. Also the suggestion made in
\cite{gnx03} to include thermal masses as kinematical masses in
decay and scattering processes leads to an unconventional picture at
temperatures $T > M_1$, which differs qualitatively from the
situation at temperatures $T < M_1$. If thermal corrections are
only included as propagator effects \cite{crx98} their influence is
small. This issue remains to be clarified. Fortunately, both
effects are only important in the case of weak washout, i.e. for
$\mt < m_*$, where the final baryon asymmetry is strongly
dependent on initial conditions in any case. In the strong washout regime,
$\mt > m_*$, which appears to be favored by the present evidence for neutrino
masses, they do not affect the final baryon asymmetry significantly.
In the following we will therefore ignore gauge boson processes and thermal
corrections. These questions will be addressed elsewhere.

The main results of this paper are summarized in the figures~6, 9 and 10. Fig.~6
illustrates that for the basic processes of decays and inverse decays the analytical
approximation for the efficiency factor agrees well with the numerical result.
The figure also
demonstrates that scatterings lead to a departure from this basic picture only for
values $K=\mt/m_* < 1$, where the final baryon asymmetry depends strongly on the initial
conditions. Fig.~9 shows the dependence of the efficiency factor on initial conditions
and on $\D L=1$ scatterings for different values of the effective Higgs mass $M_h$.
Again, for $\mt > m_*$ this dependence is small and, within the theoretical
uncertainties,
the efficiency factor is given by the simple power law
\bea
\k_f = (2 \pm 1) \times 10^{-2} \left({0.01\,{\rm eV}\over \mt}\right)^{1.1\pm 0.1}\;.
\eea
Knowing the efficiency factor, one obtains from Eqs.~(\ref{etaB}) and (\ref{cpmax})
the maximal baryon asymmetry.
Fig.~10 shows the lower bound on the initial temperature $T_{\rm i}$ as
function of $\mt$. In the most interesting mass range favored by neutrino oscillations
it is about one order of magnitude smaller than the lower bound on $M_1$. The smallest
temperature $T_{\rm i}^{\rm min} \simeq 3\times 10^9$~GeV is reached at
$\mt \simeq 2\times 10^{-3}$~eV. In sect.~4 an analytic expression for the light
neutrino mass bound is derived, which explicitly shows the dependence on the involved
parameters.

\section{Decays and inverse decays}

It is very instructive to consider first a simplified picture in which decays and
inverse decays are the only processes. For consistency, also the real intermediate state
contribution to the $2\rightarrow 2$ processes has to be included. The kinetic equations
(\ref{lg1}) and (\ref{lg2}) then reduce to
\begin{eqnarray}
{dN_{N_1}\over dz} & = & -D\ (N_{N_1}-N_{N_1}^{\rm eq}) \;, \label{dlg1} \\
{dN_{B-L}\over dz} & = &
-\varepsilon_1\ D\ (N_{N_1}-N_{N_1}^{\rm eq})-W_{I\!D}\,N_{B-L} \;,\label{dlg2}
\end{eqnarray}
where $W_{I\!D}$ is the contribution to the washout term due to inverse decays.
From Eqs.~(\ref{ef}) and (\ref{dlg1}) one obtains for the efficiency factor,
\begin{equation}\label{dlg3}
\k(z)= - {4\over 3} \int_{z_{\rm i}}^{z}\, dz'\
{dN_{N_1}\over dz'}\
e^{-\int_{z'}^{z}\ dz''\, W_{I\!D}(z'')}\;.
\end{equation}
As we shall see, decays and inverse  decays are sufficient to describe qualitatively
many properties of the full problem.

After a discussion of several useful analytic approximations we will study in detail
the two regimes of weak and strong washout. The insight into the dynamics of the
non-equilibrium process gained from the investigation of these limiting cases will then
allow us to obtain analytic interpolation formulae which describe rather accurately
the entire parameter range. All results will be compared with numerical solutions of
the kinetic equations.

\subsection{Analytic approximations}

Let us first recall some basic definitions and formulae.
The decay rate takes the form \cite{kw80},
\begin{equation}
\Gamma_D(z) = \tilde{\Gamma}_D \,
\left\langle {1\over \gamma} \right\rangle \;,
\end{equation}
where the thermally averaged dilation factor is given by the ratio of
the modified Bessel functions $K_1$ and $K_2$,
\begin{equation}\label{G}
\left\langle {1\over \gamma} \right\rangle =
{K_1(z)\over K_2(z)}\;,
\end{equation}
and $\widetilde{\Gamma}_D$ is the decay width,
\begin{equation}\label{dr}
\widetilde{\Gamma}_D={\mt M_1^2\over 8 \pi v^2}\;,
\end{equation}
with the Higgs vacuum expectation value $v=174$~GeV.
The decay term $D$ is conveniently written in the form \cite{kt}
\begin{equation}\label{D}
D(z) = K\,z\,\left\langle {1\over \gamma} \right\rangle\;.
\end{equation}

The inverse decay rate is related to the decay rate by
\begin{equation}
\Gamma_{I\!D}(z) =\Gamma_D(z)\,{N_{N_1}^{\rm eq}(z)\over N_l^{\rm eq}} \; ,
\end{equation}
where $N_l^{\rm eq}$ is the equilibrium density of lepton doublets. Since the number of
degrees of freedom for heavy Majorana neutrinos and lepton doublets is the same,
$g_{N_1}=g_l=2$, one has
\begin{equation}\label{Neq}
N_{N_1}^{\rm eq}(z) = {3\over 8} z^2 K_2(z) \; , \quad
N_l^{\rm eq} = {3\over 4}\;.
\end{equation}
The contribution of inverse decays to the washout term $W$ is therefore
\bea\label{WID1}
W_{I\!D}(z) = {1\over 2}\ {\Gamma_{I\!D}(z)\over H(z)\,z}=
{1\over 4} K z^3 K_1(z)\;,
\eea
which, together with Eqs.~(\ref{G}), (\ref{D}) and (\ref{Neq}), implies
\bea\label{WID2}
W_{I\!D}(z) = {1\over 2} D(z)\,{N_{N_1}^{\rm eq}(z)\over N_l^{\rm eq}}\;.
\eea
\begin{figure}
\centerline{\psfig{file=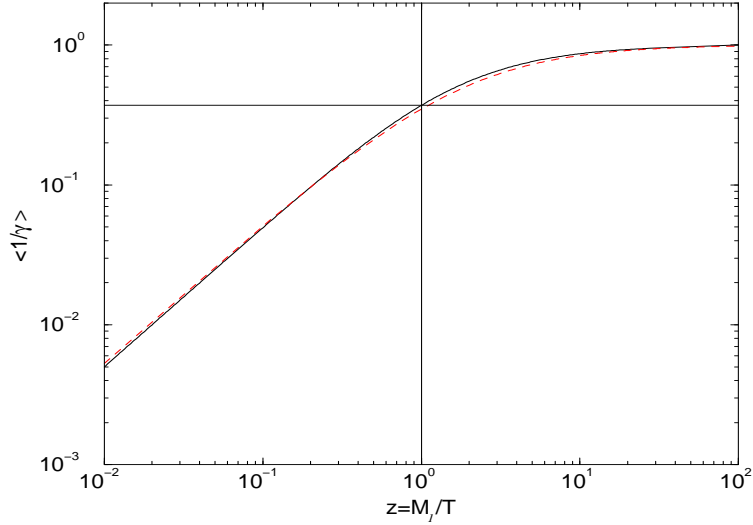,height=7cm,width=10cm}}
\caption{\small The dilation factor. The dashed line is the analytical
expression Eq.~(\ref{Dapprox}) to be compared with
the numerical result (solid line).}\label{dil}
\end{figure}

All relevant quantities are given in terms of the Bessel functions $K_1$ and $K_2$,
whose asymptotic limits are well known. At high temperatures one has,
\begin{equation}
K_2(z) \simeq {2\over z}\ K_1(z) \simeq {2\over z^2}\;, \quad z\ll 1\;,
\end{equation}
whereas at low temperatures,
\begin{eqnarray}
K_2(z) &\simeq& {1\over z}\left({15\over 8}+z \right) K_1(z) \NO\\
&\simeq& {1\over z^2} \left({15\over 8}+z \right) \sqrt{{\pi\over 2}z}\ e^{-z}\;,
\quad  z\gg 1\;.
\end{eqnarray}
Accurate interpolating functions for $K_1(z)$ and $K_2(z)$ for all values of $z$ are
\begin{eqnarray}\label{approx}
K_2(z) &\simeq& {1\over z}\left({15\over 8}+z\right) K_1(z) \NO\\
&\simeq& {1\over z^2}\left({15\over 8}+z\right) \sqrt{1+{\pi \over 2}z }\ e^{-z}\;.
\end{eqnarray}
Note, that for $z\ll 1$ this approximation gives $K_2(z) \simeq 15/(8z^2)$ rather than
the exact asymptotic form $2/z^2$. However, the high temperature domain is not so
important for baryogenesis and the approximation (\ref{approx}) is rather precise
in the more relevant regime around $z\simeq 1$.

Eq.~(\ref{approx}) yields very simple expressions for the dilation factor and the decay
term,
\begin{equation}\label{Dapprox}
\left\langle {1\over\gamma}\right\rangle (z) \simeq {z \over {15\over 8} + z} \; ,
\quad D(z) \simeq K\ {z^2 \over {15\over 8} + z}\;.
\end{equation}
As Fig.~\ref{dil} shows these analytical approximations are rather precise.
The relative error is always less than $7\%$.
The washout term (\ref{WID1}) becomes in the approximation (\ref{approx}),
\begin{equation}\label{WID}
W_{I\!D}(z) \simeq {1\over 4} K z^2 \sqrt{1+{\pi\over 2}z}\ e^{-z}\;.
\end{equation}

It is useful to define a value $z_{\rm d}$, corresponding to a `decay temperature'
$T_{\rm d}$ below which decays are {\it in equilibrium}, by
$\G_D(z_{\rm d})/H(z_{\rm d}) = z_{\rm d} D(z_{\rm d})=2$.
The value of $z_{\rm d}$ is determined by $K$, and from Eq.~(\ref{Dapprox})
one obtains
\begin{equation}
z_{\rm d}^3 - {2\over K} \left(z_{\rm d} + {15\over 8}\right) \simeq 0\;.
\end{equation}
For $K \ll 1$, this yields $z_{\rm d} \simeq \sqrt{2/K}$, whereas
$z_{\rm d} \simeq (15/4K)^{1/3}$ for $K \gg 1$. At $K \simeq 1$ one has
$z_{\rm d} \simeq 2$.

Inverse decays are {\it in equilibrium} if $W_{I\!D}(z) \geq 1$. From Eq.~(\ref{WID})
one easily finds that $W_{I\!D}(z)$ reaches its maximal value
$W_{I\!D}(z_{\rm max})\simeq 0.3\,K$ at $z_{\rm max}\simeq 2.4$. Hence, for
$K > 3$, there exists an interval $z_{\rm in}\leq z_{\rm max} \leq z_{\rm out}$,
where inverse decays are {\it in equilibrium}. For $K \lesssim 3$ no such interval exists
and inverse decays are always {\it out of equilibrium}.

\begin{figure}[t]
\centerline{\psfig{file=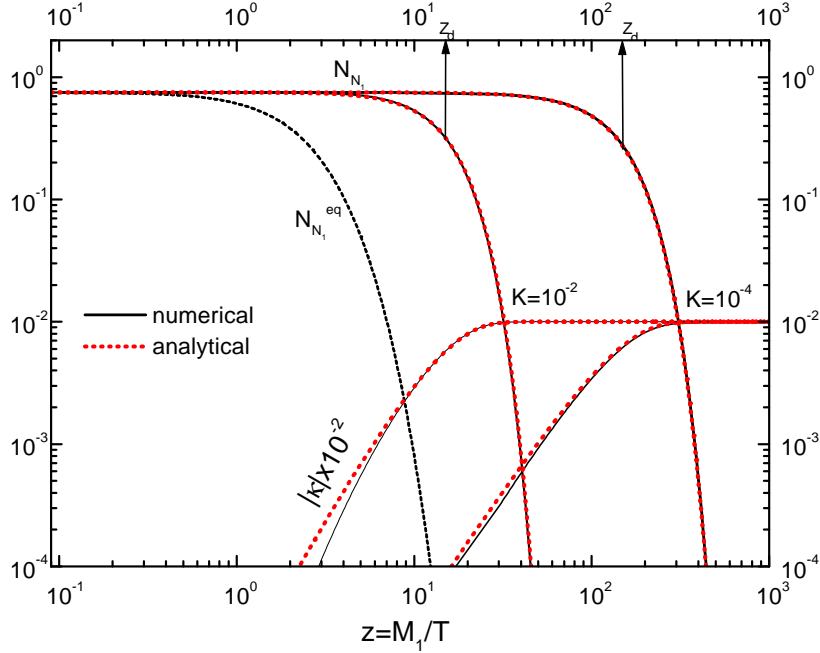,width=14cm}}
\caption{Out of equilibrium decays. $N_1$ number density, efficiency factor and decay
temperature $T_{\rm d} = M_1/z_{\rm d}$ for $K=10^{-2}$ and $K=10^{-4}$.} \label{out}
\end{figure}

\subsection{Out-of-equilibrium decays}

In the regime far out of equilibrium, $K \ll 1$, decays occur at very small
temperatures, $z_{\rm d}\gg 1$, and the produced $B-L$ asymmetry is not reduced
by washout effects. In this case the integral for the efficiency factor (\ref{dlg3})
becomes simply,
\begin{equation}\label{oodec}
\kappa(z)\simeq{4\over 3}\left(N_{N_1}^{\rm i} - N_{N_1}(z)\right) \; .
\end{equation}

For $z < z_{\rm d}$ no asymmetry is generated because the heavy neutrinos do not decay.
They also cannot be produced since inverse decays are switched off as well. Hence,
in this regime the dynamics is completely frozen. For $z > z_{\rm d}$ the equilibrium
abundance is negligible, and from Eq.~(\ref{dlg1}) one finds,
\begin{eqnarray}
N_{N_1}(z) &\simeq& N_{N_1}^{\rm i}\, e^{-\int_{z_{\rm i}}^{z}dz'\ D(z')} \NO\\
&\simeq& N_{N_1}^{\rm i}\
e^{-K\ \left({z^2 \over 2} - {15z \over 8}
+ \left({15\over 8}\right)^2\ln{\left(1+{8\over 15}z\right)}\right)} \;.
\end{eqnarray}
Note that we have neglected the small neutrino abundance
which for $N_{N_1}^{\rm i} \ll N_{N_1}^{\rm eq}$
is produced before the neutrinos decay.
Fig.~\ref{out}  shows the evolution of $N_{N_1}(z)$ and $N_{B-L}(z)$ for
$K = 10^{-2}$ and $K = 10^{-4}$, with $N_{N_1}^{\rm i}=N_{N_1}^{\rm eq}=3/4$, comparing
the numerical solutions with the analytical expressions.

The final value of the efficiency factor $\k_{\rm f} = \k(\infty)$ is proportional
to the initial $N_1$ abundance. If $N_{1}^{\rm i}=N_{1}^{\rm eq}=3/4$,
then $\k_{\rm f}=1$.
But if the initial abundance is zero, then $\k_{\rm f}=0$ as well. Therefore in
this region there is the well known problem that one has to invoke
some external mechanism to produce the initial abundance of neutrinos.
Moreover the assumption that the initial asymmetry is washed out
does not hold. Thus in the regime $K \ll 1$ the results strongly depend on
the initial conditions and the picture is not self-contained.

\subsection{Dynamical initial abundance}

In order to obtain the efficiency factor in the case of vanishing initial
$N_1$-abundance, $N_{N_1}(z_{\rm i}) \equiv N_{N_1}^{\rm i} \simeq 0$, one has to
calculate how heavy neutrinos are dynamically produced by inverse decays.
This requires solving the kinetic equation (\ref{dlg1}) with the initial condition
$N_{N_1}^{\rm i} = 0$.

Let us define a value $z_{\rm eq}$ by the condition
\begin{equation}\label{condeq}
N_{N_1}(z_{\rm eq})=N_{N_1}^{\rm eq}(z_{\rm eq}) \;.
\end{equation}
Eq.~(\ref{dlg1}) implies that the number density reaches its maximum at $z=z_{\rm eq}$.
An approximate solution can be found by noting that for $z<z_{\rm eq}$ inverse decays
dominate and thus
\begin{equation}\label{low}
{{dN_{N_1} \over dz}} \simeq D\ N_{N_1}^{\rm eq} > 0 \; .
\end{equation}
A straightforward integration yields for $z < z_{\rm eq}$
(cf.~(\ref{G}), (\ref{D}), (\ref{Neq})),
\begin{eqnarray}\label{Nprod}
N_{N_1}(z) & \simeq & {3\over 8} K \int_{z_{\rm i}}^{z} dz' z'^3 K_1(z') \NO\\
&=& {3\over 2} \int_{z_{\rm i}}^{z}\ dz'\ W_{I\!D}(z') \;.
\end{eqnarray}
In the case $z_{\rm i} \ll z < 1$, this implies for the number density,
\begin{equation}\label{NN1<}
N_{N_1}(z) \simeq {K\over 8}\,z^3 \; ,
\end{equation}
where the small dependence on $z_{\rm i}$ has been neglected.

We can now calculate the corresponding approximate solution for the
efficiency factor $\k(z)$. For $z < z_{\rm eq}$ the efficiency factor
$\k \equiv \k^-$ is negative since, $N_{N_1} < N_{N_1}^{\rm eq}$ . From
Eqs.~(\ref{dlg3}) and (\ref{low}) one obtains
\begin{eqnarray}
\k^-(z) &\simeq& -{4\over 3} \int_{z_{\rm i}}^{z} dz' D(z') N_{N_1}^{\rm eq}(z')\
e^{-\int_{z'}^{z} dz'' W_{I\!D}(z'')} \NO\\
&=& -2 \int_{z_{\rm i}}^{z} dz' W_{I\!D}(z')
e^{-\int_{z'}^{z} dz'' W_{I\!D}(z'')} \NO\\
&=& -2 \left(1-e^{-\int_{z_{\rm i}}^{z}\ dz'\ W_{I\!D}(z')} \right) \NO\\
&\simeq & -2 \left(1-e^{-{2\over 3}\ N_{N_1}(z)} \right)\;,
\quad z\leq z_{\rm eq}\;  .\label{kminus}
\end{eqnarray}
As expected, for $N_{N_1}\ll 1$ the efficiency factor is proportional to $N_{N_1}$,
up to corrections which correspond to washout effects. For $z > z_{\rm eq}$,
$\k^-(z)$ is reduced by washout effects.

For $z\geq z_{\rm eq}$, there is an additional positive contribution to the efficiency
factor,
\begin{equation}
\k^+(z) \simeq {4\over 3} \int_{z_{\rm eq}}^{z} dz' D(z')
\left(N_{N_1}(z') - N_{N_1}^{\rm eq}(z')\right)\
 e^{-\int_{z'}^{z} dz'' W_{I\!D}(z'')}\;.
\end {equation}
The total efficiency factor is the sum of both contributions,
\begin{equation}
\k_{\rm f}(z) = \k^+(z) + \k^-(z) \;.
\end{equation}
For $z\geq z_{\rm eq}$ we now have to distinguish two
different situations, the weak and strong washout regimes, respectively.

\subsubsection{Weak washout regime}

Consider first the case of weak washout, $K \ll 1$, which implies $z_{\rm eq} \gg 1$.
From Eq.~(\ref{Nprod}) one then finds,
\begin{equation}\label{nkweak}
N_{N_1}(z_{\rm eq})\simeq {9\pi \over 16}\ K \equiv N(K)\; .
\end{equation}
A solution for $N_{N_1}(z)$, valid for any $z$, is obtained by using in
Eq.~(\ref{Nprod}) the useful approximation
\begin{equation}\label{z2I1}
\int_{0}^{z}\,dz'\,z'^3\,K_1(z')\simeq {3\pi\,z^3\over [(9\pi)^{c}
+ (2\,z^3)^{c}]^{1/c}}\, ,
\end{equation}
with $c=0.7$. This yields an interpolation of the two asymptotic regimes
(cf.~Eqs.~(\ref{NN1<}) and (\ref{nkweak})), which is
in excellent agreement with the numerical result, as shown in Fig.~3a
for $K=10^{-2}$.

For $z>z_{\rm eq}\gg 1$ decays dominate over inverse decays, such that
\begin{equation}\label{high}
{{dN_{N_1} \over dz}} = - D\ N_{N_1} < 0 \; .
\end{equation}
In this way one easily obtains
\begin{equation}
N_{N_1}(z)=N_{N_1}^{\rm eq}(z_{\rm eq})
\ e^{-\int_{z_{\rm eq}}^{z} dz'\ D(z')} \;.
\end{equation}
Moreover, for $z > z^{\rm eq}$, $W_{I\!D}(z)$ is exponentially suppressed and washout
effects can be neglected in first approximation.
For the negative part of the efficiency factor one then has (cf.~Eq.~(\ref{kminus}))
\begin{equation}\label{kmweak}
\k^-(z) =  -2 \left(1-e^{-{2\over 3} N(K)} \right)\;.
\end{equation}
From Eq.~(\ref{oodec}) one obtains for the positive contribution,
\begin{equation}
\k^+(z) = {4\over 3}\left(N(K) - N_{N_1}(z)\right) \;.
\end{equation}
The final efficiency factor is then given by
\begin{equation}\label{kfprod}
\k_{\rm f}(K) \simeq {4\over 3}\ N(K) - 2 \left(1 - e^{-{2\over 3} N(K)}\right) \;.
\end{equation}

To first order in $N(K) \propto K$ the final efficiency factor vanishes.
This corresponds
to the approximation where washout effects are completely neglected. As discussed above,
$\k_{\rm f}$ is then proportional to $N_{N_1}^{\rm i}$ and therefore zero. To obtain
a non-zero asymmetry the washout in the period $z < z_{\rm eq}$ is crucial. It reduces
the absolute value of the negative contribution $\k^{-}(z)$, yielding a positive
efficiency factor of order ${\cal O}(K^2)$,
\begin{equation}\label{reduction}
\k_{\rm f}(K) = \left[{2\over 3}\,N(K)\right]^2
\simeq {9\pi^2\over 64}\,K^2 \; .
\end{equation}
Such a reduction of the generated asymmetry has previously been observed in the
context of GUT baryogenesis \cite{ft81}.
Note, that for $K > 1$ Eq.~(\ref{reduction}) does not hold, since in this case
$z_{\rm eq}$ becomes small and washout effects for $z \geq z_{\rm eq}$ are also
important.

\begin{figure}
\centerline{\psfig{file=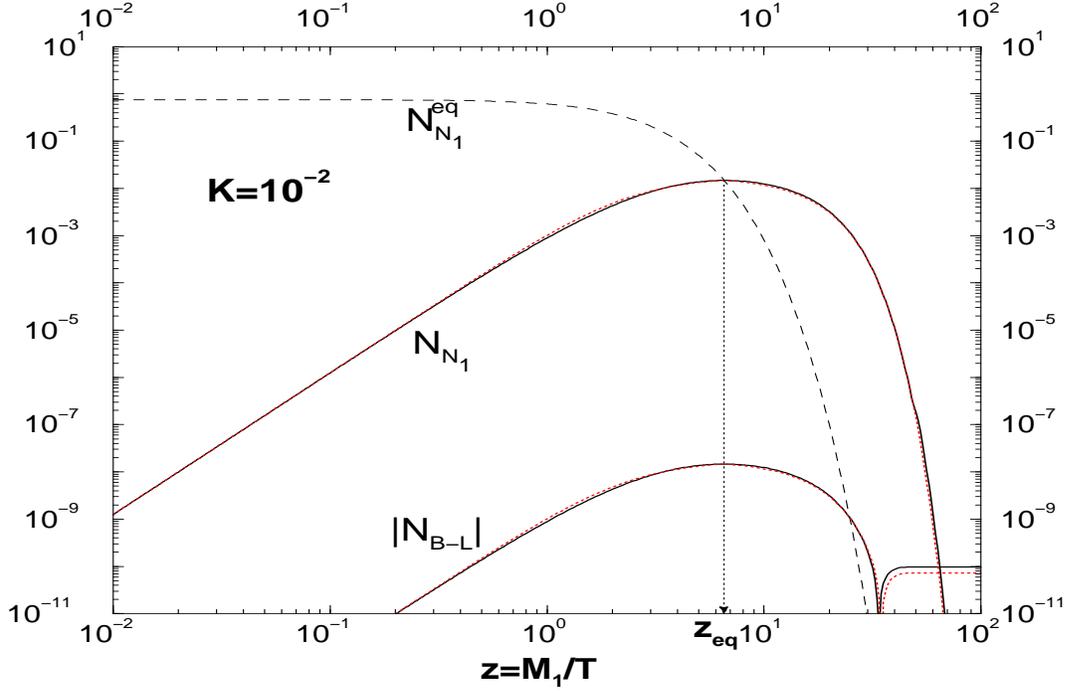,height=9cm,width=14cm}}
\vspace{20mm}
\centerline{\psfig{file=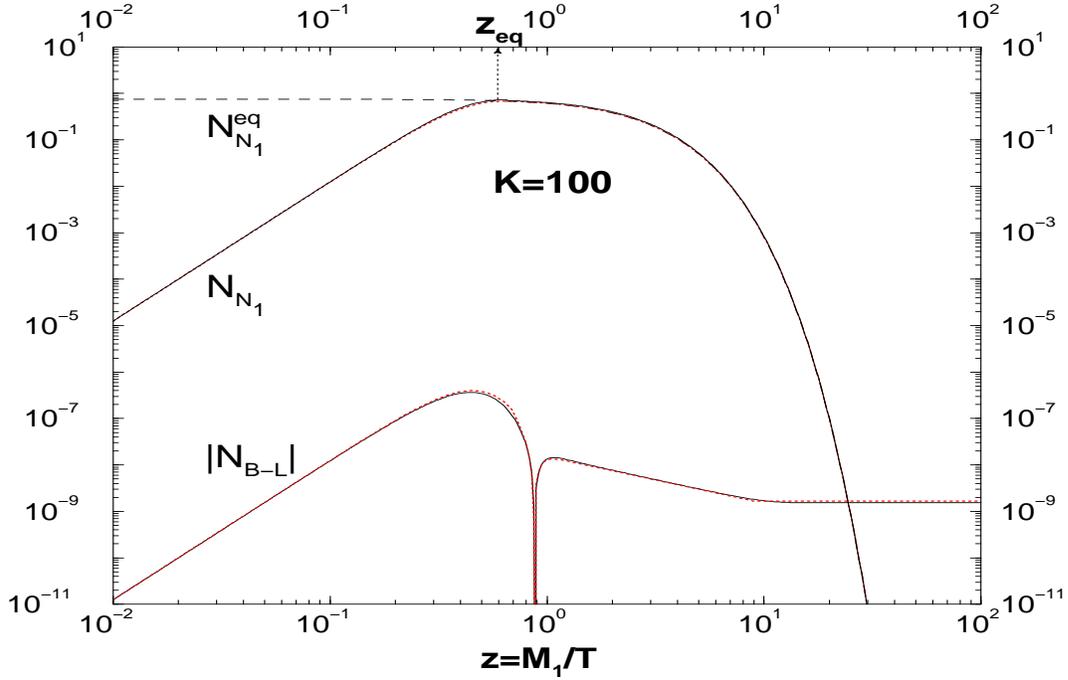,height=9cm,width=14cm}}
\caption{Comparison analytical (dashed lines) and numerical (solid lines) results  for
heavy neutrino production and $B-L$ asymmetry in the case of zero initial abundance,
$N_{N_1}^{\rm i}=0$, for weak washout (top) and strong washout (bottom); $|\ve_1|=10^{-6}$.}
\label{prod}
\end{figure}

In Fig.~\ref{prod}a the analytical solutions for $N_{N_1}(z)$ and
$|N_{B-L}(z)| = (3/4)|\ve_1\k(z)|$ are compared with the numerical results for
$K\simeq 10^{-2}$. A residual asymmetry survives after
$z_{\rm d} \gg z_{\rm eq}$ as remnant of the cancellation between
the negative and the positive contributions to the efficiency factor.
The second one is prevalent because washout suppresses $\kappa^-$ more efficiently.
As one can see in Fig.~\ref{prod}a,
the analytical solution for the asymmetry slightly underestimates the final
numerical value. This is because for $K\gtrsim 10^{-2}$ the approximation
of neglecting washout for $z\geq z_{\rm eq}$ becomes inaccurate.

\subsubsection{Strong washout regime}

As $K$ increases, $z_{\rm eq}$ decreases, and at $K \simeq 3$ the maximal number density
$N(K)$ reaches the equilibrium density $N_{eq}$ at $z_{\rm eq}\simeq 1$.
For $K\gg 1$, one obtains from Eq.~(\ref{NN1<}),
$z_{\rm eq} \simeq (6/K)^{1/3}\simeq z_{\rm d}\ll 1$. A more accurate description
for $z\lesssim z_{\rm eq}$ has to take into account decays in addition to inverse
decays, i.e. one has to solve Eq.~(\ref{dlg1}).
Since $z_{\rm eq}\ll 1$, one can use $N_{N_1}^{\rm eq}\simeq 3/4$, and one
then easily finds,
\begin{equation}
N_{N_1}(z)={3\over 4}\,\left(1-e^{-{1\over 6}K\,z^3}\right),
\end{equation}
which correctly reproduces Eq.~(\ref{NN1<}) for $z\ll z_{\rm eq}$.
An example, with $K=100$, is shown in Fig.~\ref{prod}b which illustrates
how well the analytical expression for neutrino production agrees
with the numerical result.

Consider now the efficiency factor. For $K\gg 1$ we can neglect the negative
contribution $\k^-$, assuming that the asymmetry generated at high temperatures
is efficiently washed out.
This is practically equivalent to assuming thermal initial abundance. We will see
in the next section how to describe the transition from the weak to the
strong washout regime.

For $K\gtrsim 3$, inverse decays are in equilibrium in the range
$z_{\rm in} < z < z_{\rm out}$, with $z_{\rm in} \simeq 2/\sqrt{K}$. In the strong
washout regime, $K \gg 1$, the efficiency factor can again be calculated analytically.

For $z\lesssim z_{\rm d} \simeq [15/(4\,K)]^{1/3}$ decays are not effective in tracking
the equilibrium distribution. For the difference
\begin{equation}
\D = N_{N_1}(z)-N_{N_1}^{\rm eq}(z),
\end{equation}
with $N_{N_1}(0)=N_{N_1}^{\rm eq}(0)=3/4 \equiv N_{\rm eq}$, one  has,
\begin{equation}\label{del1}
\Delta(z)\simeq
N_{N_1}^{\rm i} - N_{N_1}^{\rm eq}(z) \simeq {3\over 16}\ z^2\;,
\quad z\lesssim z_{\rm d}\; .
\end{equation}
The corresponding efficiency factor is given by (cf.~(\ref{ef}))
\begin{equation}\label{kzsmall}
\k(z)\simeq {4\over 3} \int_{z_{\rm i}}^{z}
dz' D(z') \D(z') \simeq {2 K\over 75} z^{5}\;, \quad  z\lesssim z_{\rm d}\;.
\end{equation}

On the other hand, for $z > z_{\rm d}$ the neutrino abundance tracks
closely the equilibrium behavior. Since $D \propto K$, one can solve Eq.~(\ref{dlg1})
systematically in powers of $1/K$, which yields
\begin{equation}\label{del}
\D(z) = -{1\over D} {dN_{N_1}^{\rm eq}\over dz} + {\cal O}\left({1\over K^2}\right)\;.
\end{equation}
Using the properties of Bessel functions, Eq.~(\ref{Neq}) yields for the
derivative of the equilibrium density,
\begin{equation}\label{dNeq}
{{dN_{N_1}^{\rm eq} \over dz}} = -{3\over 8} z^2 K_1(z) = -
{3\over 2Kz} W_{I\!D}(z)\; .
\end{equation}

We can now calculate the efficiency factor. From Eqs.~(\ref{dlg3}) and (\ref{dNeq})
one obtains
\begin{eqnarray}\label{efd}
\k(z) &=&
{2\over K} \int_{z_{\rm i}}^{z} dz'\, {1\over z'} W_{I\!D}(z')\,
e^{- \int_{z'}^{z} dz''\, W_{I\!D}(z'')} \NO\\
&\equiv& \int_{z_{\rm i}}^{z} dz'\, e^{- \j(z',z)} \;.
\end{eqnarray}
The integral is dominated by the contribution from a region around the value $\bar{z}$
where $\j(z',z)$ has a minimum. The condition for a local minimum $z_{\rm B}$, the
vanishing of the first derivative, yields
\begin{equation}\label{eqz0}
W_{I\!D}(z_{\rm B}) = \left\langle {1\over\gamma}\right\rangle^{-1}(z_{\rm B})\,
-\, {3\over z_{\rm B}}\;.
\end{equation}
Since the second derivative at $z_{\rm B}$ is positive one has
$\bar{z}={\rm min}\{z,z_{\rm B}\}$.

The integral (\ref{efd}) can be evaluated systematically by the steepest descent method
(cf.~\cite{kt}).
Alternatively, a simple and very useful approximate analytical solution can be obtained
by replacing in the exponent of the integrand $W_{I\!D}(z)$ by
\begin{equation}
\overline{W}_{I\!D}(z) = {\bar{z}\over z}\, W_{I\!D}(z)
= - {K\bar{z}\over 4}\, {d\over dz} (z^2 K_2(z))  \;.
\end{equation}
The efficiency factor then becomes
\begin{eqnarray}\label{kz}
\k(z) &\simeq& {2\over K\bar{z}} \int_{z_{\rm i}}^{z} dz'\, \overline{W}_{I\!D}(z')
\, e^{-\int_{z'}^{z} dz''\, \overline{W}_{I\!D}(z'')} \NO\\
& = & {2\over K \bar{z}}\
\left(1-e^{-\int_{z_{\rm i}}^{z} dz'\, \overline{W}_{I\!D}(z') )} \right) \;.
\end{eqnarray}
It is now easy to understand the behavior of $\k(z)$. For
$z_{\rm d}\lesssim z < z_{\rm B}$, one has $\k \propto 1/z$,
while for $z \geq z_{\rm B}$ the efficiency factor gets frozen at a final value
$\k_{\rm f}\simeq 2/(K\,z_{\rm B})$, up to a small correction ${\cal O}(\exp{(-K)})$.

\begin{figure}[t]
\centerline{\psfig{file=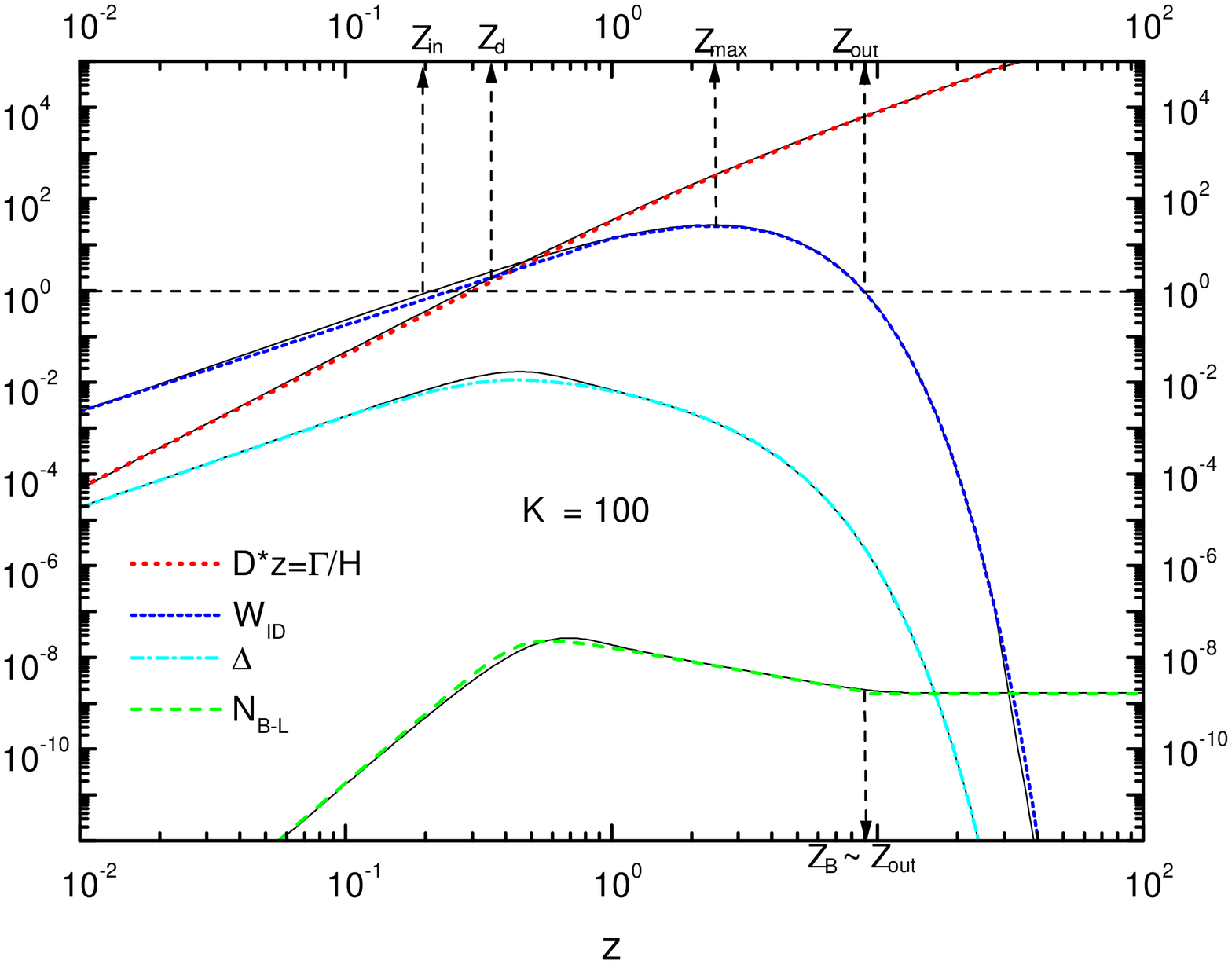,height=10cm,width=16cm}}
\caption{\small Strong washout: comparison between analytical and numerical
(full lines) results.
Inverse decays are in equilibrium in the temperature range
$z_{\rm in} \leq z \leq z_{\rm out} \sim z_B$; $|\ve_1| = 10^{-6}$.}
\label{strong}
\end{figure}

One can also easily find global solutions for all values of $z$ by
interpolating the asymptotic solutions for $z<z_{\rm d}$ and $z>z_{\rm d}$,
respectively. From Eqs.~(\ref{D}), (\ref{del1}), (\ref{del}) and (\ref{dNeq}) one obtains
for the difference between $N_1$-abundance and equilibrium abundance,
\begin{equation}\label{Delta}
\Delta(z)\simeq \left(1 + {K z^3\over {15\over 4} + 2 z}\right)^{-1}
{3\over 16} z^3 K_1(z) \; .
\end{equation}
Similarly, an interpolation between the expressions (\ref{kzsmall}) and (\ref{kz})
for the efficiency factor is given by
\begin{equation}\label{kstrong}
\k(z) \simeq \left(1 + {K^2 \bar{z} z^5\over 75}\right)^{-1}{2 K\over 75} z^5  \; .
\end{equation}

A typical example of strong washout is shown in Fig.~\ref{strong} for the value
$K=100$, as in Fig.~\ref{prod}b, but now for thermal initial abundance. In this
figure we also show the decay, inverse decay and washout terms. Instead of the
neutrino abundance the deviation $\Delta(z)$ is shown.
The dotted, short dashed, dot-dashed and dashed lines are the approximations
Eq.~(\ref{Dapprox}) for the $D$ term, Eq.~(\ref{WID}) for the $W_{I\!D}$ term,
Eq.~(\ref{Delta})
for $\Delta(z)$ and Eq.~(\ref{kstrong}) for $\k(z)$, respectively. The thin solid lines
are the numerical results which agree well with the analytical approximations.
The behavior $\k(z) \propto 1/z$ for $z > z_{\rm d}$ and the freeze-out of
$N_{B-L}$ at $z_{\rm out}$ are clearly visible.

Let us now focus on the final value of the efficiency factor $\k_{\rm f} = \k(\infty)$.
Note, that for $K\gg 1$ also $z_{\rm B}\gg 1$, and the condition (\ref{eqz0}) becomes
approximately $W_{I\!D}(z_{\rm B})\simeq 1$. This means $z_{\rm B}\simeq z_{\rm out}$.
Hence, the asymmetry produced for $z \leq z_{\rm out}$ is essentially washed out, while for
for $z > z_{\rm out}$ washout is negligible ($W_{I\!D} < 1)$.
This simple picture will have some interesting consequences and applications.

The integral in Eq.~(\ref{kz}) is easily evaluated,
\begin{equation}
\int_{0}^{\infty} dz'\,\overline{W}_{I\!D}(z) = {1\over 2} z_{\rm B}(K) K \;.
\end{equation}
Using the approximations (\ref{D}) and (\ref{WID}), the condition (\ref{eqz0})
for $z_{\rm B}(K)$ becomes explicitly
\begin{equation}\label{z0a}
{K\over 4}\,z_{\rm B}(K)^{3}\,e^{-z_{\rm B}(K)} \,
\sqrt{1+{\pi\over 2}\,z_{\rm B}(K)} \simeq z_{\rm B}(K) - 1 \;.
\end{equation}
$z_B(K)$ approaches one as $K$ goes to zero\footnote{Note that the solution
$z_B(K)$ of Eq.~(\ref{eqz0}) approaches asymptotically 1.33 for $K \rightarrow 0$.
However, in the strong washout regime and also for the extrapolation $K \rightarrow 0$
this difference is irrelevant for $\k_{\rm f}$.}. For $K \gg 1$ the solution of
Eq.~(\ref{z0a}) is given by
\begin{equation}
  z_B(K) \simeq -{5\over2}\,W_{-1}\left(-{4\over5\pi^{1/5}}K^{-2/5}\right)\;,
  \label{lam}
\end{equation}
where $W_{-1}$ is one of the real branches of the Lambert W function
\cite{lambert}. This result can be approximated by using the
asymptotic expansion of $W_{-1}$ \cite{lambert,lambert2},
\begin{equation}
  z_B(K) \simeq{1\over2}\,\mbox{ln}\left({\pi K^2\over1024}
     \left[\mbox{ln}\left({3125\pi K^2\over1024}\right)\right]^5\right)
     +{\cal O}\left({\mbox{ln}(K)\over\mbox{ln}(\mbox{ln}(K))}\right)\;.
  \label{zB2}
\end{equation}
\begin{figure}[t]
\centerline{\psfig{file=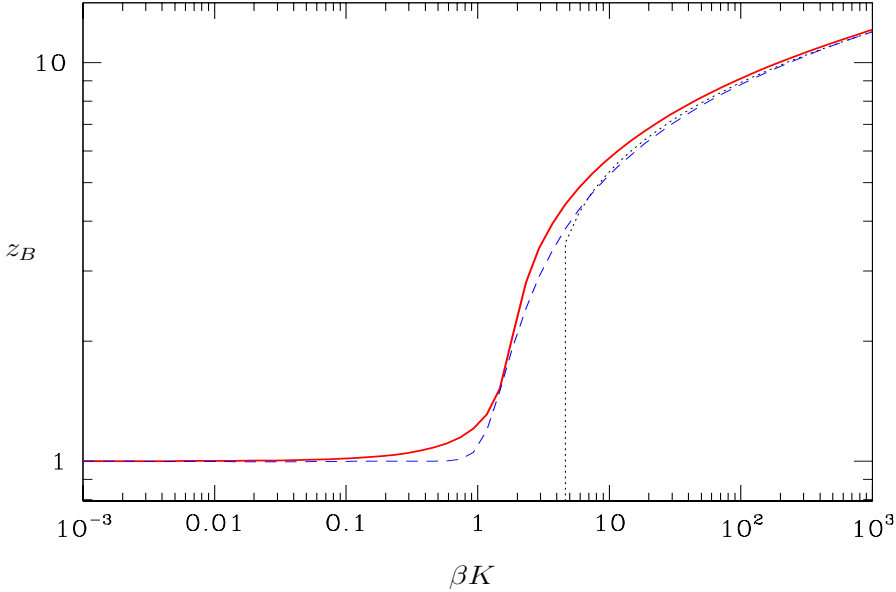,height=8cm,width=12cm}}
\caption{$z_B$ as function of the decay parameter $\beta K$.
The case studied in this section corresponds to $\beta=1$.
The solid red line is the numerical solution of eq.~(\ref{z0a}), the
dotted black line shows the asymptotic solution (\ref{lam}),
and the blue dashed line is the interpolation (\ref{interpol}).
\label{z0}}
\end{figure}
A rather accurate expression for $z_B(K)$ for all values of $K$ is given by the
interpolation (cf.~Fig.~5),
\begin{equation}
  z_B(K) \simeq 1+{1\over2}\,\mbox{ln}\left(1+{\pi K^2\over1024}
     \left[\mbox{ln}\left({3125\pi K^2\over1024}\right)\right]^5\right)\;.
  \label{interpol}
\end{equation}
Note the rapid transition from strong to weak washout at $K \simeq 3$.

The final value of the efficiency factor takes the simple form
\begin{equation}\label{kfan}
\k_{\rm f}(K) \simeq
{2\over z_{\rm B}(K)K}\left(1-e^{-{1\over 2}z_{\rm B}(K)K}\right)\;.
\end{equation}

\begin{figure}[t]
\centerline{\psfig{file=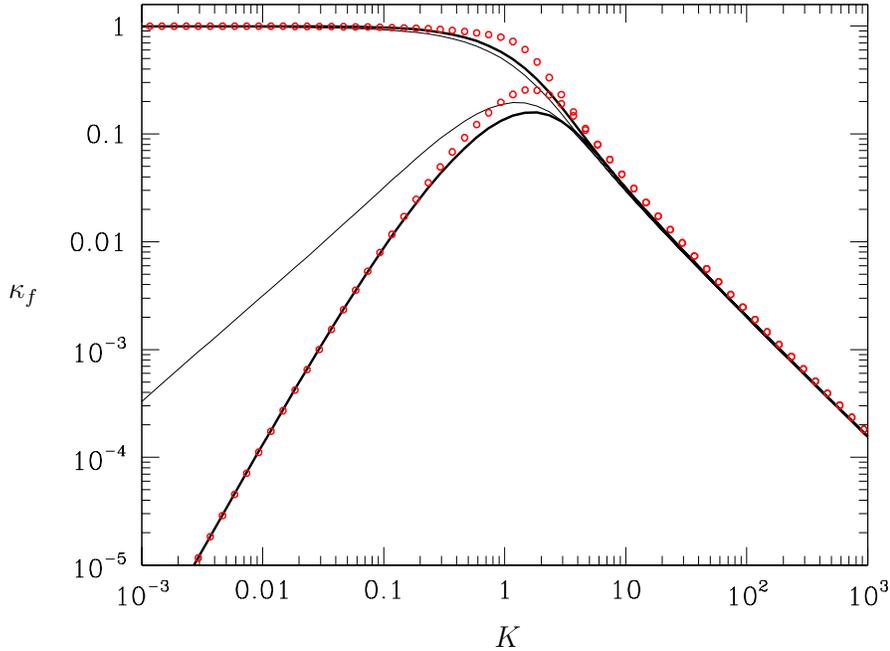,height=9cm,width=12cm}}
\caption{The final efficiency factor $\k_{\rm f}$ as function of the decay parameter
$K$ for thermal and dynamical initial $N_1$ abundance, respectively. The thick solid
lines are the numerical solutions. The thin lines show, for comparison, the numerical
solutions of the complete Boltzmann equations including $N_1$-top scatterings with
an effective Higgs mass $M_h/M_1=0.1$. The red circles represent Eq.~(\ref{kfan})
for the efficiency factor, evaluated using Eq.~(\ref{interpol}) for $z_B(K)$.}
\label{deckf}
\end{figure}

This analytical expression for the final efficiency factor, combined
with Eq.~(\ref{interpol}) for $z_B(K)$, provides an accurate description of
$\k_{\rm f}(K)$, as shown in Fig.~\ref{deckf}. Eq.~(\ref{kfan}) can also be extrapolated
into the regime of weak washout, $K \ll 1$,
where one obtains $\k_{\rm f} = 1$ corresponding to thermal initial abundance,
$N_{N_1}^{\rm i}=N_{N_1}^{\rm eq} = 3/4$. It turns out that also in the transition
region Eqs.~(\ref{kfan}) and (\ref{interpol}) provide a rather accurate description,
as is evident from Fig.~\ref{deckf}.
The largest discrepancy between
analytical and numerical results is about 30\%  around $K \sim 1$. For
comparison also the numerical result including scatterings is shown. The difference
with respect to the basic `decay-plus-inverse-decay' picture becomes significant only
for $K < 1$.

The above analysis is easily extended to the case where the strength of the washout
term $W_{I\!D}$ is modified to $\b W_{I\!D}$. For instance, in the model considered in
\cite{kt}, $B-L$ number changes by two in heavy particle decays, corresponding to
$\b =2$. On the other hand, the heavy particle abundance is not affected by this change.
The final efficiency factor is therefore given by
\begin{equation}\label{betakf}
\k_{\rm f}(K)\simeq {2\over z_B(K)\b K}\,
\left(1-e^{-{1\over 2}z_B(K)\b K} \right)\;,
\end{equation}
where $z_B(K)$ is again given by Eq.~(\ref{z0a}).
In the regime $K\gg 1$ our expression for the final efficiency factor can be
approximated by
\begin{equation}\label{kfKl}
\k_{\rm f}(K)\simeq {2\over z_B(K)\b K} \simeq {1\over
1.2\, \b K\, (\ln{\b K})^{0.8}} \;,
\end{equation}
which is  very similar to the result\footnote{ Quantitatively, a
discrepancy by a factor $\sim 7$ was noted in \cite{news}, which
is mostly related to the definition of the decay parameter $K$
(cf.~Eq.~(\ref{decpar})).} obtained by Kolb and Turner \cite{kt}.

Comparing the efficiency factor (\ref{kfan}) with the solution of the Boltzmann equations
including scatterings, as shown in Fig.~6, one arrives
at the conclusion that the simple decay-plus-inverse-decay picture represents a very
good approximation for leptogenesis in the strong washout regime. As we will see,
the difference is essentially negligible within the  current theoretical
uncertainties.

\subsubsection{Global parametrization}

Given the results of the previous sections it is straightforward to obtain an
expression for the efficiency factor for all values of $K$ also in the case of
dynamical initial abundance. We first introduce a number density $\overline{N}(K)$
which interpolates between the maximal number densities $N_{\rm eq}=3/4$ and
$N(K)=9\p K/16$ (cf.~(\ref{nkweak})) for strong and weak washout, respectively,
\begin{equation}\label{nkall}
\overline{N}(K) = {N(K)\over\left(1 + \sqrt{{N(K)\over N_{\rm eq}}}\right)^2}\; .
\end{equation}

The efficiency factor is in general the sum of a positive and a negative contribution,
\begin{eqnarray}
\k_{\rm f}(K) = \k_{\rm f}^+(K) + \k_{\rm f}^-(K) \;. \NO
\end{eqnarray}
Here $\k_{\rm f}^-$(K) is given by (\ref{kmweak}) for $K\ll 1$.
A generalization, accounting for washout also for $z\geq z_{\rm eq}$, reads
\begin{equation}\label{kmgen}
\k^-(z)=\k^-(z_{\rm eq})\,
e^{{2\over 3}N(K) - \int_0^z\,dz'\,W_{\rm ID}(z')}
\end{equation}
This expression extends the validity of the analytical solution to values
$K > 10^{-2}$ in the case of a dynamical initial abundance.
$\k_{\rm f}$ is exponentially suppressed for $K\gg 1$. An interpolation, satisfying
the asymptotic behaviors at small and large $K$, is given by
\begin{equation}\label{km}
\k^{-}_{\rm f}(K) = -2\ e^{-{2\over 3}N(K)}
\left(e^{{2\over 3} \overline{N}(K)} - 1 \right)\; .
\end{equation}

On the other hand, the expression (\ref{kfan}) for $\k_{\rm f}^+(K)$, which is valid for
$K\gg 1$, has to approach $4/3\, N(K)$ for $K\ll 1$. These requirements are fulfilled by
\begin{equation}\label{kp}
\k^{+}_{\rm f}(K)={2\over z_B(K)K}
\left(1-e^{-{2\over 3} z_B(K)K \overline{N}(K)}\right) \; .
\end{equation}
Equations (\ref{nkall}), (\ref{km}) and (\ref{kp}), together with the interpolation
(\ref{interpol}) for $z_B(K)$,
yield an accurate description of the efficiency factor $\k_{\rm f}(K)$ for all values
of $K$, as demonstrated by Fig.~\ref{deckf}.

This result is a good starting point for obtaining
an analytic description of the efficiency factor for the full problem. In the following
sections we shall go beyond the simple decay-and-inverse-decay picture and
include other processes step by step.

\section{The scattering term}

\subsection{Analytic approximations}

The scattering term $S$ and the related washout contribution $W_{\D L=1}$
arise from two different classes of Higgs and lepton mediated inelastic scatterings
involving the top quark ($t$) and gauge bosons ($A$),
\begin{equation}
S=S^{t}+S^{A}\;.
\end{equation}
Their main effect is to enhance the neutrino production and thus the efficiency
factor for $\mt < m_*$. Further, they also contribute to the washout term, which leads
to a  correction of the efficiency factor for $\mt > m_*$, i.e. in the strong washout
regime. Since the scattering processes are specific to leptogenesis we shall use in
this section mostly the variable $\mt \geq m_1$ \cite{fhy02} rather than $K$.
Top quark and gauge
boson scattering terms are expected to be of similar size. However, the reaction
densities for the gauge boson processes are presently controversial \cite{pu03,gnx03}.
We shall therefore discuss these processes in detail elsewhere. We shall also neglect
the scale dependence of the top-Yukawa coupling, which reduces the size of $S^t$,
since this decrease of $S$ will be partially compensated by $S^A$.

The term $S^{t}$ is again the sum of two terms arising from the s-channel processes
$N_1\,l \leftrightarrow t\,q$ and the t-channel processes $N_1\,t \leftrightarrow l\,q$,
$N_1\,\bar{q} \leftrightarrow l\,\bar{t}$,
\begin{equation}
S^{t} = 2\,S_{\f,s}+4\,S_{\f,t} \; .
\end{equation}
The scattering terms are defined as usual in terms of scattering rates and
expansion rate,
\bea\label{Srates}
S_{\f,s(t)} = {\G_{\f,s(t)}^{(N_1)}\over Hz}\;,
\eea
and introducing the functions $f_{\f,s(t)}(z)$ (cf.~appendix B)
it is possible to write
\bea\label{St}
S^t={K_S\over 6} \left(f_{\f,s}(z)+2\,f_{\phi,t}(z)\right)\;;
\eea
here we have introduced the ratio
\begin{equation}
K_S={\mt\over m_*^{s}} \; ,
\end{equation}
with
\begin{equation}
m_*^s ={4\p^2\over 9}{g_{N_1} v^2 \over m_t^2}\,m_* \simeq 10\, m_* \;.
\end{equation}
At high temperatures, $z \ll 1$, the functions $f_{\f,s(t)}$ have the following
asymptotic form,
\bea
f_{\phi,s}(z) &\simeq& 2
\left[1-z^2 \left(\ln\left({2\over z}\right)-\gamma_E\right)\right] \; ,\\
f_{\phi,t}(z) &\simeq& 2
\left[1+ {z^2\over 2} \ln\left({M_1\over M_h}\right)\,
\left(\ln\left({2\over z}\right)-\g_E\right)\right] \; .
\eea

\begin{figure}[t]
\vspace{30mm}
\centerline{\psfig{file=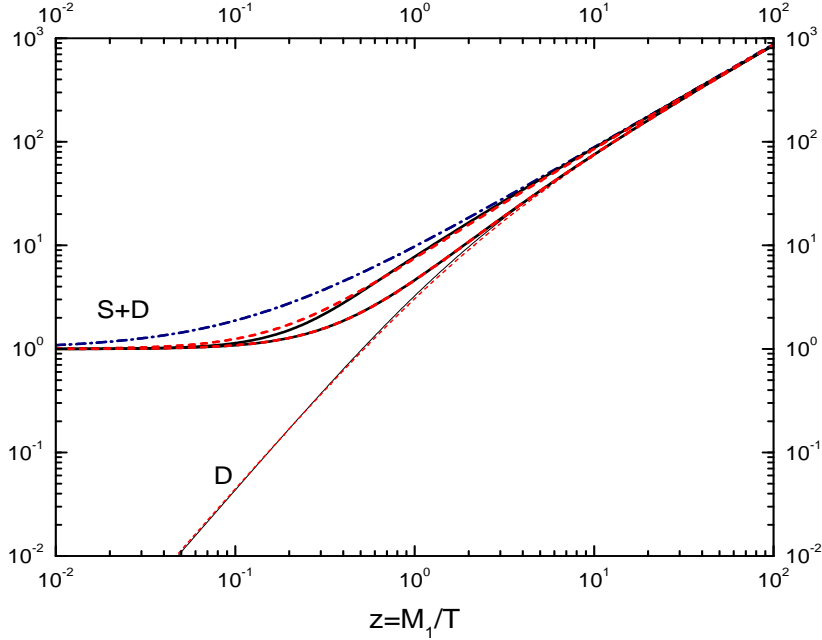,height=10cm,width=14cm}}
\vspace{-35mm}
\caption{\small The rates $S+D$ and $D$ are shown as function of $z$ for $K_S=1$, i.e.
$\mt=m_{\star}^s$. The dash-dotted line is the
simple approximation $S+D\simeq K_S+K\,z$, while the two short  dashed lines
represent Eq.~(\ref{S+D}) for $M_h/M_1=10^{-5}$ (higher) and $M_h/M_1=10^{-1}$ (lower),
to be compared with the numerical results (thick solid lines). The dotted line shows
the expression (\ref{D}) for $D$, to be compared with the numerical result for $D$
(solid line).} \label{S}
\end{figure}

In Fig.~\ref{S} we have plotted the rates $S+D$ and $D$ as function of $z$ for $K_S=1$,
i.e. $\mt = m^s_*$. For values $z<2$ the sum $S+D$ is dominated by the scattering rate
$S$, while for $z>2$ the decay rate $D\simeq K\,z$ dominates. A simple analytic
approximation for the sum $D+S$ is given by
\begin{equation}\label{S+D}
D+S \simeq K_S\,\left[1+
\ln\left(M_1\over M_h \right)\,z^2\,\ln\left({1+{a\over z}}\right) \right]\;,
\end{equation}
where
\begin{equation}
a={K\over K_S\,\ln(M_1/M_h)}= {8\,\pi^2\over 9\,\ln(M_1/M_h)}\,\,\, .
\end{equation}
Here we have introduced the Higgs mass $M_h$ to cut off the infrared divergence of the
t-channel process.
As Fig.~\ref{S} illustrates, the approximation (\ref{S+D}) agrees well with the numerical
result for $M_{h}/M_1=10^{-5}$ as well as $M_{h}/M_1=10^{-1}$. Note that the latter value
corresponds to the thermal Higgs mass, $M_h\simeq 0.4\,T$,
at the baryogenesis temperature $T_B$ in the strong washout regime.

%\vspace{20mm}

The washout term induced by the $\D L=1$ scattering processes is again the sum of
s- and t-channel contributions,
\begin{equation}
W_{\D L=1} = W_{\f,s} + 2\,W_{\f,t} \;.
\end{equation}
The washout rates are directly related to the scattering rates (\ref{Srates}),
\bea
W_{\f,t} &=& {\G^{l}_{\f,t}\over H\,z} =
{N_{N_1}^{\rm eq}\over N_l^{\rm eq}}\, S_{\f,t}\;, \\
W_{\f,s} &=& {N_{N_1}\over N_{N_1}^{\rm eq}}{\G^{l}_{\f,s}\over H\,z} =
{N_{N_1}^{\rm eq}\over N_l^{\rm eq}}\,{N_{N_1}\over N_{N_1}^{\rm eq}}\, S_{\f,s}\;.
\eea
Using Eq.~(\ref{WID2}) for the equilibrium number densities one obtains
\bea
W_{\D L=1} = 2 W_{I\!D}{1\over D}
\left({N_{N_1}\over N_{N_1}^{\rm eq}}\, S_{\f,s} + 2 S_{\f,t}\right)\;.
\eea
The washout rate including inverse decays is then given by
\bea\label{w0}
W_0 &=& W_{I\!D}+W_{\D L=1} \NO\\
&=& W_{I\!D}\left(1 +
{1\over D}\left(2 {N_{N_1}\over N_{N_1}^{\rm eq}}\,S_{\f,s} + 4 S_{\f,t}\right)\right)\;.
\eea
This is the total washout rate as long as $\D W$, the off-shell contribution from
heavy neutrinos, can be neglected. This is
justified for sufficiently small values of $M_1$ (cf.~sect.~4).

\subsection{Dynamical initial abundance}

We can now calculate the production of heavy neutrinos
and study how the efficiency factor gets enhanced by the presence of the
scattering term. We again define a value $z_{\rm eq}$ by the condition (\ref{condeq}),
\bea
N_{N_1}(z_{\rm eq})=N_{N_1}^{\rm eq}(z_{\rm eq}) \;. \NO
\eea
For $z < z_{\rm eq}$ the number density can be obtained by integrating the equation
\begin{equation}\label{dN<}
{{dN_{N_1}\over dz}} \simeq (D+S)\,N_{N_1}^{\rm eq}\,> 0 \;.
\end{equation}
The result is given by the expression ($z\leq z_{\rm eq}$)
\begin{equation}\label{N<}
N_{N_1}(z)={3\over 8}\,K_S\,\left[{\cal I}_A(z)
+\ln\left({M_1\over M_h}\right)\,{\cal I}_B(z)\,\right] \;.
\end{equation}
Here the first integral is given by
\begin{equation}\label{IA}
{\cal I}_A(z)=\int_0^z\,dz'\,z'^2\, K_2(z')\simeq {3\pi\,z^3\over
[(9\pi)^{c} + (2\,z^3)^{c}]^{1/c}} + z^3\,K_2(z)\; ,
\end{equation}
where we have used the approximation Eq. (\ref{z2I1}).
The second integral can be expressed as
\bea\label{IB}
{\cal I}_B(z) &=& \int_0^{z}\,dz'\,z'^4\,\ln\left(1+{a\over z'}\right)\,K_2(z') \NO\\
&\simeq& 2 \int_0^1\,dz'\,z'^2\,\ln\left(1+{a\over z'}\right) +
a\,\int_1^{z}\,dz'\,z'^3\,K_2(z') \NO\\
&\simeq& {2\over 3} \left((1+a^3)\ln(1+a) - a^3 \ln{a} - a^2
+{1\over 2} a\right) +a\,K_3(1)- a\,z^3\,K_3(z) \; . \eea
The value $z_{\rm eq}$ can now be determined by setting $N_{N_1}$, as
determined from Eqs.~(\ref{N<}), (\ref{IA}) and (\ref{IB}), equal
to $N_{N_1}^{\rm eq}$. Using an approximate form for $K_3$ , one obtains an equation
similar to Eq.~(\ref{z0a}), as described in appendix~B. This yields a good approximation
for $z_{\rm eq}$ in the case $\mt < m_*$.

\subsubsection{Weak washout regime}

Consider now the case of weak washout, $\mt \ll m_* \simeq 10^{-3}\ {\rm eV}$, which
implies $z_{\rm eq} \gg 1$. For $z > z_{\rm eq}$, decays dominate over inverse decays,
\bea\label{dN>}
{{dN_{N_1}\over dz}} \simeq -(D+S)\,N_{N_1}< 0 \;.
\eea
Using $D+S \simeq K z$, valid for $z \gg a$ (cf. (\ref{S+D}) and ~fig.~\ref{S}),
this yields for the number density the simple expression
\bea\label{N>}
N_{N_1}(z) &\simeq& N_{N_1}^{\rm eq}(z_{\rm eq})
\,e^{-\int_{z_{\rm eq}}^{z}\,dz'\,(S+D)} \NO\\
&\simeq& N_{N_1}^{\rm eq}(z_{\rm eq})\,e^{-{K\over 2}(z^2-z^2_{\rm eq})}.
\eea

\begin{figure}[t]\label{figSprod}
\vspace{30mm}
\centerline{\psfig{file=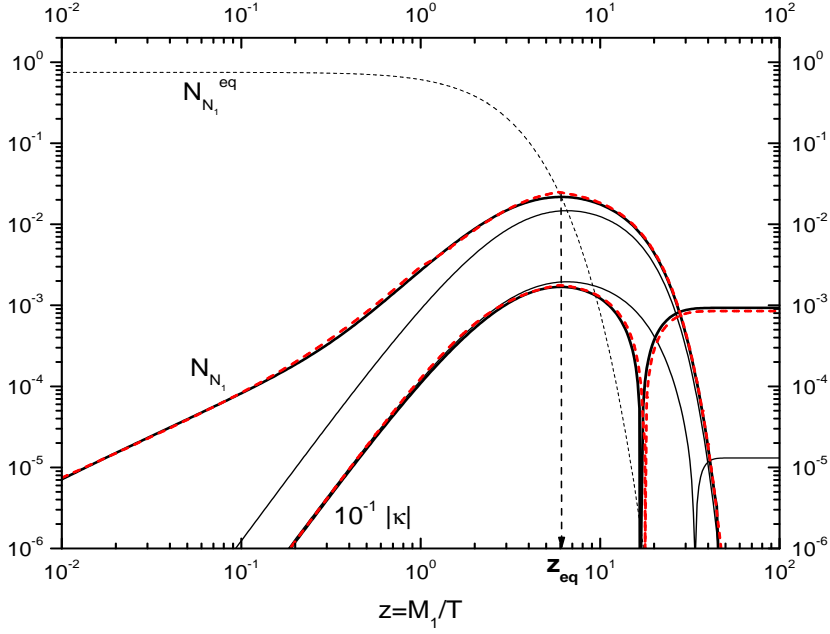,height=10cm,width=14cm}}
\vspace{-35mm}
\caption{\small Effect of scatterings on neutrino production for
$\mt=10^{-5}\,{\rm eV}$.
The numerical results have been calculated for $\Delta W=0$ and
$M_h/M_1=10^{-5}$ (thick solid lines). The short-dashed lines are the analytical
solutions for $N_{N_1}$ (Eqs.~(\ref{N<}) and (\ref{N>})) and
for $\k(z)$ (Eqs.~(\ref{k<}) and (\ref{k>})), with $M_h/M_1=10^{-5}$.
For comparison we also show the result where scatterings are neglected (thin solid
lines).}
\label{prods}
\end{figure}

In Fig.~\ref{prods} the solution $N_{N_1}(z)$ is shown for $\mt=10^{-5}\,{\rm eV}$.
The analytical solution agrees well with the numerical result.
We also make a comparison with
the result already displayed in Fig.~\ref{prod}, where the S term
is neglected. As expected, the presence of the S term enhances the density $N_{N_1}$
at $z=z_{\rm eq}$. Moreover the comparison illustrates the strong sensitivity
of the efficiency factor in the case $\mt \ll m_*$, not only to the initial
conditions, but also to the theoretical description.
A difference in $N_1$ abundance by less than a factor of two
at $z_{\rm eq}$ corresponds to final efficiency factors which differ by two orders of
magnitude. This is due to delicate cancellations between the positive and the negative
contribution to the efficiency factor and is a source of large theoretical uncertainties
in the small $\mt$ regime.

We can now calculate the efficiency factor. In sect.~2.3.1 we have seen that in the
absence of scatterings the inclusion of the small washout term was necessary to create
an asymmetry between the negative contribution, $\k^{-}$, and the positive one,
$\k^+$, in order to have a non-zero final value $\k_f$. Given the $S$ term one can
neglect washout to first approximation. From Eq.~(\ref{ef}) one then obtains
\begin{equation}\label{efS}
\k(z)= -{4\over 3}\int_{z_{\rm i}}^{z}\,dz'\,
j^{-1}\ {dN_{N_1}\over dz'} \; ,
\end{equation}
where (cf.~(\ref{D}), (\ref{S+D}))
\bea
j(z) = {D+S\over D} \simeq
\left[{z\over a}\,\ln\left(1+{a\over z}\right)+{K_S\over K\,z}\right]\,
\left(1+{15\over 8\,z}\right)\;.
\eea
Due to the S term we now have $\k_{\rm f}\neq {4\over 3} N_{N_1}^{\rm i}$,
although washout is neglected. The reason is quite clear: as in the case without
scatterings, the asymmetry is changed only by decays and inverse decays; however,
the number of decaying neutrinos at $z_{\rm eq}$ is now larger because of the
additional production due to scatterings. To first approximation
we can thus calculate the efficiency factor neglecting washout.

For $z \leq z_{\rm eq}$ one obtains (cf.~(\ref{z2I1})),
\bea\label{k<}
\k^-(z)=
-{4\over 3}\int_{z_{\rm i}}^{z}\,dz'\,D\,N_{N_1}^{\rm eq}
%=-{K\over 2}\,\int_{z_{\rm in}}^{z}\,dz'\,z'^2\,{\cal I}_1(z')
\simeq -{3\pi\,K\,z^3\over [(18\,\pi)^c + (4z^{3})^c]^{1/c}} \;.
\eea
For $z>z_{\rm eq}$ one has $j \simeq {\rm const.}$ (cf.~Fig.~\ref{S});
Eq.~(\ref{efS}) the yields the simple result,
\bea\label{k>}
\k(z) = {4\over 3}\,\left(N_{N_1}(z_{\rm eq})\,j^{-1}(z_{\rm eq})
- N_{N_1}(z)\,j^{-1}(z)\right)\,+ \,\k^-(z_{\rm eq})\;,
\eea
which is shown in Fig.~\ref{prods} for $\mt=10^{-5}$ and $M_h/M_1=10^{-5}$ (short dashed line);
it agrees reasonably well with the corresponding numerical solution (solid line). The
analytical solution somewhat overestimates $|\k^{-}_{\rm f}|$; correspondingly,
the final value $\k_f$ is underestimated.

The analytical solution explains why the final value of the efficiency factor,
$\k_{\rm f}$, is proportional to $\mt$,
\bea\label{Skfnow}
\k_{\rm f}\simeq
{4\over 3}\left[N_{N_1}(z_{\rm eq})\,j^{-1}(z_{\rm eq})
- \widetilde{N}(K)\right]
\,\propto \,\mt\;.
\eea
\begin{figure}[t]
\vspace{30mm}
\centerline{\psfig{file=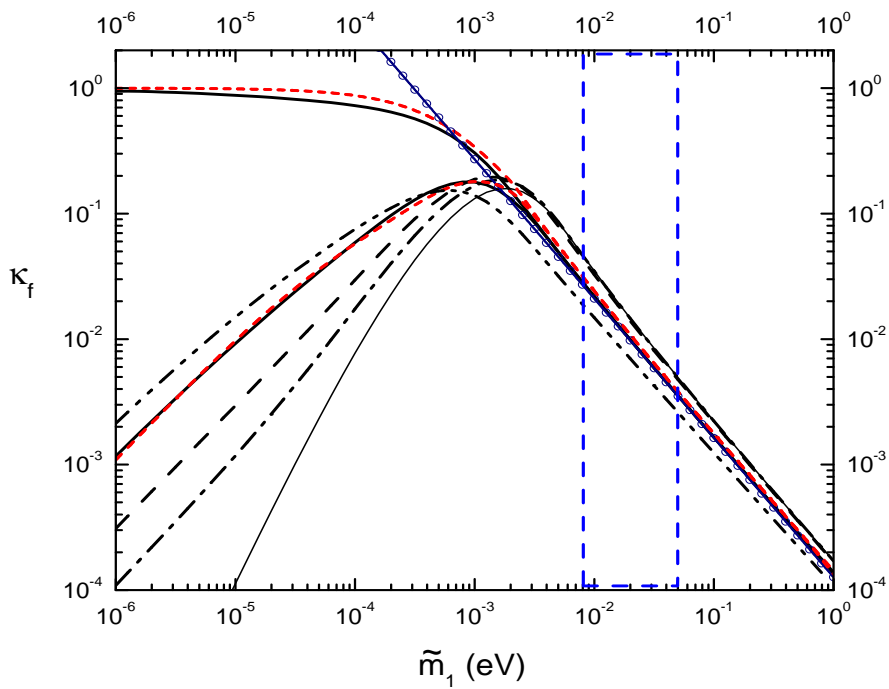,height=10cm,width=14cm}}
\vspace{-35mm}
\caption{\small The final efficiency factor when scatterings are included.
The numerical results are shown for $M_h/M_1=10^{-10},10^{-5},10^{-1},1$
(dot-dot-dash, solid,
dashed and dot-dashed line respectively). The thin dashed
line is the simple result from the decay-plus-inverse-decay picture when
scatterings are neglected. The short-dashed lines are the analytical results
in the case of thermal initial abundance (Eq.~(\ref{Skfstrong}) with
$j(z_B)\rightarrow j^2(z_B)$ in the exponential)
and zero initial abundance  (Eqs.~(\ref{Skf+}) + (\ref{Skf-}), $M_h/M_1=10^{-5}$).
The circled line is the power law fit (\ref{powerlaw}).
The dashed box indicates the range ($m_{\rm sol}$,$m_{\rm atm}$).}
\label{Skf}
\end{figure}
where
\bea\label{NtildeK}
\widetilde{N}(K) =
{2N(K)z^3_{\rm eq} \over \left((9\p)^c + (2z^3_{\rm eq})^c\right)^{1/c}} \; ,
\eea
with $N(K)=9\p K/16$. Contrary to the case discussed in sect.~2.3.1, for which
$N_{N_1}(z_{\rm eq})\simeq \widetilde{N}(K) \simeq \overline{N}(K)$ and $j=1$,
$N_{N_1}(z_{\rm eq})$ and $\widetilde{N}(K)$ are now different. Hence, there is no
cancellation of terms ${\cal O}(\mt)$ between $\k^+$ and $\k^-$.

The expression (\ref{Skfnow}) for the final efficiency factor
fails for effective neutrino masses $\mt > 10^{-5}\,{\rm eV}$. Including washout
mainly reduces the negative contribution $\k^-$ and thereby enhances the final value of
the efficiency factor. Eq.~(\ref{k<}) is then changed into
\bea\label{kfmeno}
\k^{-}_{\rm f} = -{4\over 3}\,\int_{z_{\rm i}}^{\infty}\,dz'\,D N_{N_1}^{\rm eq}\,
e^{-\int_{z'}^{\infty}\,dz''\,W_0(z'')}\;.
\eea
For $z<z_{\rm eq}$ one has $N_{N_1} < N_{N_1}^{\rm eq}$, and the washout rate
becomes (cf.~(\ref{w0}))
\bea
W_0(z)\simeq W_{I\!D}(z){D+4\,S_{\f,t}\over D} \;.
\eea
From the expression (\ref{S+D}) for $D+S$ one obtains
\bea
D+4\,S_{\f,t} &\simeq&
K_S\,\left({2\over 3}+
\ln\left(M_1\over M_h \right)\,z^2\,\ln\left(1+{a\over z}\right) \right) \NO\\
&\simeq& {2\over 3}\,K_s+K\,z \; ;
\eea
here the last approximation requires $a\ll 1$. Together with Eq.~(\ref{D}) this yields
\bea
W_0(z)\simeq W_{I\!D}(z)\left(1 + {\a\over z}\right)\;,
\eea
where the coefficient $\a$ is given by
\bea
\a = {2K_S\over 3K} + {15\over 8}\;.
\eea
Since $W_{I\!D}/z$ is a total derivative, one obtains for the efficiency factor,
\bea
\k^{-}_{\rm f} &=& -2\,
\int_{z_{\rm i}}^{\infty}\,dz'\,W_{ID}(z')\,e^{-{1\over 4} K\a {z'}^2 K_2(z')}\,
e^{-\int_{z'}^{\infty}\,dz''\,W_{I\!D}(z'')} \NO\\
&\simeq& -2\,e^{-{1\over 2}K\a}\,
\int_{z_{\rm i}}^{\infty}\,dz'\,W_{ID}(z')\,
e^{-\int_{z'}^{\infty}\,dz''\,W_{I\!D}(z'')} \;,
\eea
where we have used $z^2 K_2(z)\simeq 2$ for $z\lesssim 1$. Except for the
exponential pre-factor, this yields the result obtained in sect.~2.3.1 for decays and
inverse decays (cf.~(\ref{kmweak})) when the wash-out at $z>z_{\rm eq}$ is neglected
\bea\label{wwws}
\k^{-}_{\rm f} = -2\,e^{-{1\over 2}K\a}\,\left(1-e^{-{2\over 3}N(K)}\right)\;.
\eea
The case without scatterings is recovered for $\a=0$.

\subsubsection{Strong washout regime}

In the case of strong washout, $K\gg 1$, the density of heavy neutrinos follows
closely the equilibrium abundance, and  one can obtain an analytical solution
repeating the discussion in sect.~2.3.2. The efficiency factor is now given by
\bea\label{efSW}
\k(z)= - {4\over 3}\,\int_{z_{\rm in}}^{z}\,dz'\,j^{-1}\,
{dN_{N_1}^{\rm eq}\over dz'}\,e^{-\int_{z'}^z\,dz''\,W_0(z'')} \;.
\eea
Using $N_{N_1}/N_{N_1}^{\rm eq} \simeq 1$ in the washout rate one obtains
(cf.~(\ref{w0})),
\bea\label{W0}
W_0 \simeq W_{I\!D}\,j\;.
\eea
In this way one finds for the efficiency factor
\bea\label{psis}
\k(z) &=& {2\over K}\,\int_{z_{\rm i}}^{z}\,dz'\,{1\over z'j(z')}\,
W_{I\!D}(z')\,e^{- \int_{z'}^z\,dz''\,W_{I\!D}(z'') j(z'')} \NO\\
&\equiv& \int_{z_{\rm i}}^{z}\,dz'\,e^{-\j_S(z',z)}\;.
\eea

As in sect.~2.3.2, the dominant contribution to the integral
arises from a region around a value $z_B \gg 1$ where $\psi_S(z',z)$
has a minimum. Since $D + S \simeq K_S + K z$ for large $z$, the value $z_B$ is again
given by  Eq.~(\ref{eqz0}) up to corrections ${\cal O}(K_S/(K\,z_0^2))$. Replacing now
$W_{I\!D}(z)$ by $W_{I\!D}(z)z_B j(z_B)/(zj(z))$ in  the exponent of the integrand
and by $W_{I\!D}(z) j(z)/j(z_B)$ in the pre-factor, respectively, one obtains for the
final efficiency factor the approximate solution,
\begin{equation}\label{Skfstrong}
\k_{\rm f}={2\over z_B K \,j(z_B)^2}
\left(1-e^{- {1\over 2} z_B K j(z_B)} \right) \;.
\end{equation}
This extends Eq.~(\ref{betakf}) to the case where scatterings are included.

Note, that at small $K$ the efficiency factor ({\ref{Skfstrong}}) does not approach
one, the value corresponding to thermal initial abundance. However, any initial abundance
can be reproduced by adjusting the exponent in Eq.~({\ref{Skfstrong}}). Replacing $j(z_B)$
by $j(z_B)^2$ leaves Eq.~(\ref{Skfstrong}) essentially unchanged at large $K$, whereas
at small $K$ one has $\k_{\rm f} \simeq 1$ corresponding to thermal initial abundance.
The result is shown in Fig.~\ref{Skf} (short-dashed line) and compared
with numerical results for different values of $M_{\rm h}/M_1$. In the strong
washout regime, $\mt \gg m_*$, and for $M_h/M_1=10^{-5}$, the analytical and numerical
results agree within $10\%$.
Since the strong washout regime is most interesting with respect to neutrino mass
models, this is one of the most relevant results of this paper.

For practical purposes it is interesting to note that, within the current theoretical
uncertainties, the efficiency factor for $\mt > m_*$ is given by the simple power law,
\begin{equation}\label{powerlaw}
\kappa_{\rm f}\simeq (2\pm 1) \times 10^{-2}\,
\left({0.01\ {\rm eV}\over \mt}\right)^{1.1\pm 0.1} \; .
\end{equation}
The quoted uncertainties represent, approximately, the range visible in Fig.~9 for large $\mt$.
It is limited from above by the thin solid line corresponding to decays plus inverse
decays and from below by the dot-dot-dashed line where scatterings are included with
the extremely small ratio $M_h/M_1=10^{-10}$. Note that $M_h/M_1=0.1$ corresponds
to a thermal Higgs mass, $M_h\simeq 0.4\,T$, at the
baryogenesis temperature $T_B \simeq M_1/5$.

\subsubsection{Global parametrization}

As in sect.~2.3.3 we can now obtain an expression for the efficiency factor for
all values of $K$ by interpolating between the two regimes of small $K$ and
large $K$. We shall use the number density $\widetilde{N}(K)$
(cf.~(\ref{NtildeK})) and the interpolation (\ref{interpol}) for $z_B(K)$,
which is related to the baryogenesis temperature by $T_B = M_1/z_B$.

The efficiency factor is the sum of a positive and a negative contribution,
\bea
\k_{\rm f}(K) = \k_{\rm f}^+(K) + \k_{\rm f}^-(K) \;. \NO
\eea
Here $\k_{\rm f}^-(K)$ differs from Eq.~(\ref{km}) just by the exponential pre-factor
induced by the scatterings (cf.~(\ref{wwws})), which yields
\begin{equation}\label{Skf-}
\k^{-}_{\rm f}(K) = -2\ e^{-{2\over 3}\left(N(K)+{3\over 4}\,K\a\right)}\,
\left(e^{{2\over 3} \widetilde{N}(K)} - 1 \right)\; .
\end{equation}
The expression (\ref{Skfstrong}) for $\k_{\rm f}^+(K)$, which is valid for
$K\gg 1$, has to approach Eq.~(\ref{Skfnow}) for $K\ll 1$. An interpolating function is
given by
\begin{equation}\label{Skf+}
\k_{\rm f}^{+}(K)={2\over z_B(K) K\,j^2(z_B)}\,
\left(1-e^{- {2\over 3}\,z_B(K) K
 j^2(z_B)N_{N_1}(z_{\rm eq})\,j^{-1}(z_{\rm eq})}\right) \, .
\end{equation}
Eq.~(\ref{kp}) is recovered for $j=1$ and $S=0$. The sum
$\k_{\rm f}=\k_{\rm f}^{+}+\k_{\rm f}^{-}$ is shown in Fig.~\ref{Skf}
for $M_h/M_1=10^{-5}$. The agreement with the numerical result is very good.
Including washout yields a description which correctly interpolates between the
weak washout regime, $\mt \ll m_*$ and the strong washout regime, $\mt \gg m_*$.

\subsection{Lower bounds on $M_1$} \label{M1}

The results for the efficiency factor are easily translated into
theoretical predictions for the observed baryon-to-photon ratio
using the relation (\ref{etaB}). The theoretical prediction has to
be compared with the results from WMAP \cite{WMAP} combined with
the Sloan Digital Sky Survey \cite{SDSS}, $\O_b\,h^2=0.023\pm
0.001$, corresponding to
\begin{equation}\label{etanew}
\eta_B^{CMB} = (6.3\pm 0.3)\times 10^{-10}  \; .
\end{equation}
The comparison yields the required $C\!P$ asymmetry in terms of the
baryon-to-photon ratio and the efficiency factor $\k_{\rm f}$ (cf.~(\ref{etaB})),
\begin{equation}
\ve_1^{CMB}={\eta_B^{CMB}\over d\,\kappa_{\rm f}}\simeq 6.3\times
10^{-8} \left(\eta_B^{CMB}\over 6\times 10^{-10}\right)\,\k_{\rm f}^{-1}\;.
\end{equation}

The $C\!P$ asymmetry $\ve_1$ can be written as product of a
maximal asymmetry and an effective leptogenesis phase $\d_{L}$ \cite{hmy02},
\begin{equation}
\ve_1=\ve_1^{\rm max}\, \sin\,\d_L \;.
\end{equation}
The connection between the leptogenesis phase and other $C\!P$ violating observables is
an important topic of current research \cite{phase03}.
The maximal $C\!P$ asymmetry $\ve_1^{\rm max}$ depends in general on $M_1$,
$\mt$ and, via the light neutrino masses $m_i$, on the absolute neutrino mass scale
$\mb$ \cite{bdp02}.
For given light neutrino masses, i.e. fixed $m_1$ and $m_3$,
$\ve_1$ is maximized in the limit $m_1/\mt \rightarrow 0$,
for which one obtains \cite{di02},
\begin{equation}
\ve_1^{\rm max}(M_1,\mb)= {3\over 16\p}{M_1\over v^2}\left(m_3-m_1 \right)\;.
\end{equation}
This expression reaches its maximum for fully hierarchical
neutrinos, corresponding to $m_1=0$ and $m_3=m_{\rm atm}\equiv
\sqrt{\D m^2_{\rm atm}}$.

Neutrino oscillation experiments give for atmospheric neutrinos \cite{bari,atm}
\begin{equation}
\D m^2_{\rm atm}= (2.6\pm 0.4)\times 10^{-3}\,{\rm eV}^2\;,
\end{equation}
and for solar neutrinos \cite{solar}
\begin{equation}
\D m^2_{\rm sol}\simeq  (7.1^{+1.2}_{-0.6} \times 10^{-5})\,{\rm
eV^2}\;,
\end{equation}
implying
\begin{equation}\label{matm}
m_{\rm atm}=(0.051\pm 0.004)\,{\rm eV} \, .
\end{equation}

Thus, apart the small experimental error, $m_{\rm atm}$ is a fixed
parameter and the resulting maximal asymmetry depends uniquely on
$M_1$ \cite{di02},
\begin{equation}\label{cpmax}
\ve_1^{\rm max}(M_1)={3\over 16\p}{M_1\,m_{\rm atm}\over v^2}
\simeq 10^{-6}\,\left(M_1\over 10^{10}\,{\rm GeV} \right)
\,\left({m_{\rm atm}\over 0.05\,{\rm eV}}\right)\,\;.
\end{equation}

The maximal $C\!P$ asymmetry, or equivalently the maximal
leptogenesis phase, corresponds to a maximal baryon asymmetry
$\eta_B^{\rm max}$. The CMB constraint $\eta_B^{\rm
max}\geq\eta_B^{\rm CMB}$, together with Eq.~(\ref{cpmax}), then
yields a lower bound on the heavy neutrino mass $M_1$,
\bea\label{lbM1} M_1 > M_1^{\rm min} &=& {1\over d}\,{16\,\pi\over
3}\,{v^2\over m_{\rm atm}}\,
{\eta_B^{CMB}\over\k_{\rm f}} \NO\\
&\simeq& 6.4\times 10^{8}\,{\rm GeV} \left(\eta_B^{CMB}\over
6\times 10^{-10}\right) \left(0.05\,{\rm eV}\over m_{\rm
atm}\right)\,\k_{\rm f}^{-1} \;. \eea

Note that the bound depends on the combination
$\eta_B^{CMB}/m_{\rm atm}$ whose error, after the WMAP result,
receives a similar contribution both from $\eta_B^{CMB}$ and $m_{\rm atm}$ such that
\begin{equation}\label{lowM1}
M_1^{\rm min}(\mt)=(6.6\pm 0.8)\times 10^8\,{\rm GeV}\,\k_{\rm f}^{-1}(\mt)
\gtrsim 4\times 10^8\,{\rm GeV}\,\kappa_{\rm f}^{-1}(\mt) \; ,
\end{equation}
with the last inequality indicating the $3\sigma$ lower bound.
For values of $M_1 \ll 10^{14}\,{\rm GeV}$ \cite{bdp02} one can
use the results of this section for the efficiency factor, neglecting
the $\D L=2$ washout term. Eq.~(\ref{lbM1})
then provides a lower bound on $M_1$ which depends only on $\mt$.
\begin{figure}[t]
\centerline{\psfig{file=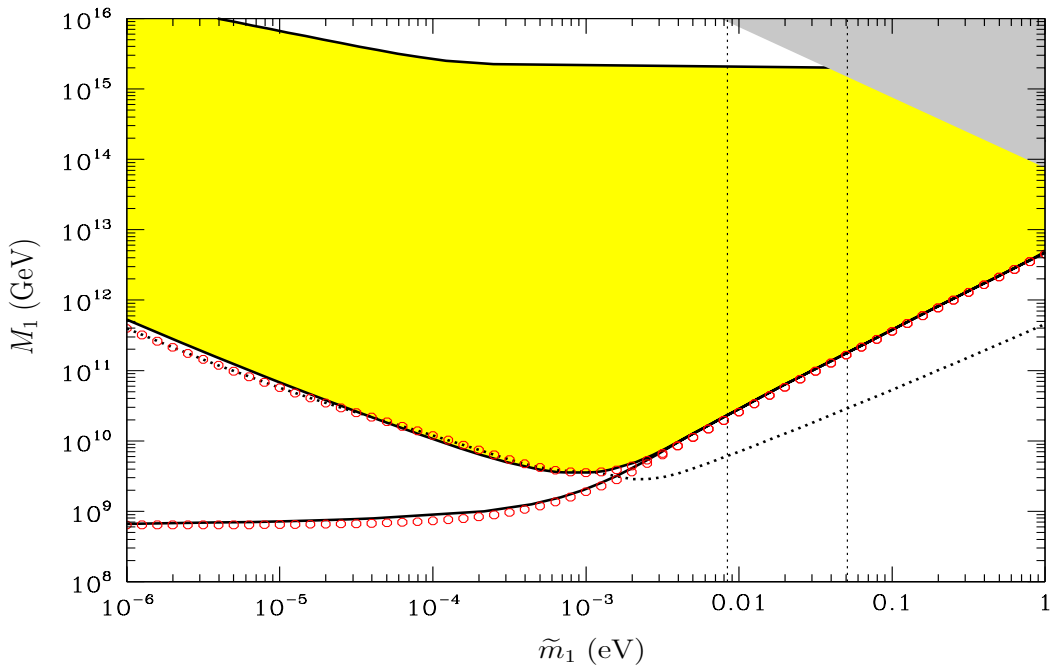,height=9cm,width=14cm}}
\caption{Analytical lower bounds on $M_1$ (circles) and $T_{\rm i}$ (dotted line) for
$m_1 = 0$, $\eta_B^{CMB} = 6\times 10^{-10}$ and $m_{\rm atm} = 0.05\,{\rm eV}$.
The analytical results are compared with the numerical ones (solid lines).
The vertical dashed lines indicate the range ($m_{\rm sol}$,$m_{\rm atm}$).
The gray triangle at large $M_1$ and large $\mt$ is excluded by theoretical
consistency (cf.~appendix~A).}
\label{figbounds}
\end{figure}

Fig.~\ref{figbounds} shows the analytical results for $M_1^{\rm min}(\mt)$, based on
Eq.~(\ref{Skfstrong}) for thermal initial
abundance (thin lines) and the sum of Eqs.~(\ref{Skf-}) and
(\ref{Skf+}) for zero initial abundance (thick lines). For
comparison also the numerical results (solid lines) are shown.
The absolute minimum for
$M_1$ is obtained for {\it thermal initial abundance} in the limit
$\mt\rightarrow 0$, for which $\k_{\rm f}=1$. The corresponding
lower bound on $M_1$ can be read off from Eq.~(\ref{lowM1})
and at $3\,\sigma$ one finds
\begin{equation}\label{M1lbth}
M_1 \gtrsim 4\times 10^8\,{\rm GeV} \;.
\end{equation}
This result is in agreement with \cite{bdp02} and also with the recent
calculation \cite{gnx03}. Note that the lower bound on $M_1$ becomes much more
stringent in the case of only two heavy Majorana neutrinos \cite{ct03}.
The bound for thermal initial abundance
is model independent. However, it relies on some unspecified
mechanism which thermalizes the heavy neutrinos $N_1$ before the
temperature drops considerably below $M_1$. Further, the case $\mt
\ll 10^{-3}\,{\rm eV}$ is rather artificial within neutrino mass
models, and in this regime a pre-existing asymmetry would not be
washed out \cite{bdp03}.

For {\em zero initial abundance} the lower bound is obtained for
$\k_0(\mt)=\k_{\rm peak}\simeq 0.18$, corresponding to $\mt^{\rm
peak}\simeq 10^{-3}\,{\rm eV}$. In this case one obtains from
Eq.~(\ref{lowM1}) \cite{bdp02},
\begin{equation}\label{M1lbza}
M_1 \gtrsim 2 \times 10^{9}\,{\rm GeV} \; .
\end{equation}
Particularly interesting is the lower bound on $M_1$ in the
favored neutrino mass range $m_{\rm sol} \lesssim \mt \lesssim
m_{\rm atm}$. This  range lies in the strong washout
regime where the simple power law scaling (\ref{powerlaw}) holds.
One thus obtains from Eq.~(\ref{lowM1})
\begin{equation}
M_1^{\rm min}(\widetilde{m}_1)\simeq (3.3\pm 0.4) \times
10^{10}\,{\rm GeV}\, \left(\widetilde{m}_1\over 10^{-2}\,{\rm eV}
\right)^{1.1} \,\, ,
\end{equation}
which at $3\,\sigma$ implies
\begin{equation}\label{M1lbfr}
M_1\gtrsim 2\times 10^{10}\,{\rm GeV}\, \left(\widetilde{m}_1\over
10^{-2}\,{\rm eV} \right)^{1.1} \simeq (10^{10} \div 10^{11})\,{\rm GeV} \; ,
\end{equation}
where the last range corresponds to values $m_{\rm sol}\lesssim
\widetilde{m}_1\lesssim m_{\rm atm}$. Note that these bounds are
fully consistent with neglecting the $\D L=2$ washout term, which
is justified for $M_1^{\rm min} \ll 10^{14}\,{\rm GeV}$.

In the case of near mass degeneracy between the lightest and the
next-to-lightest heavy neutrino $N_2$, $y = (M_2 - M_1)/M_1 \ll 1$, the upper bound
on the $C\!P$ asymmetry $\ve_1$ is enhanced by a factor $\xi(y) \simeq 1/(3 y)$
\cite{fps95,crv96}. The $C\!P$ asymmetry of $N_2$ can also be maximal. Since the number of
decaying neutrinos is essentially doubled, one obtains for the
reduced lower bound on $M_1$ ($y < 1$),
\begin{equation}
M_1 > M_1^{\rm min}(\mt,y) = {M_1^{\rm min}(\mt)\over 2\,\xi(y)} \; ,
\end{equation}
where $M_1^{\rm min}$ is given by Eq.~(\ref{lbM1}). It has been
suggested that for extreme degeneracies a resonant regime
\cite{pu03} is reached where $\ve_1^{\rm max} ={\cal O}(1)$. In this
case there is practically no lower bound on $M_1$ from leptogenesis.

\subsection{Lower bound on $T_{\rm i}$}

It is usually assumed that the lower bound on the initial temperature $T_{\rm i}$
roughly coincides with $M_1^{\rm min}$, the lower bound on the heavy neutrino mass
$M_1$. Here $T_{\rm i}$ can be thought of as the temperature after reheating, below
which the universe is radiation dominated \cite{gkr01}. However, in the following we
will show that this is only the case in the weak washout regime, i.e. for
$\mt \lesssim m_* \simeq 10^{-3}\,{\rm eV}$, whereas in the more interesting strong washout
regime $T_{\rm i}$ can be about one order of magnitude smaller than $M_1$.

In general, the maximal baryon asymmetry is a function of both, $M_1$ and
$T_{\rm i}=M_1/z_{\rm i}$,
\begin{equation}
\eta_B^{\rm max}(M_1,\mt,\mb,z_{\rm i})=
d\,\ve_1^{\rm max}(M_1,\mt,\mb)\,
\k_{\rm f}(\mt,z_{\rm i}) \; .
\end{equation}
In a rigorous procedure one would have to treat $M_1$ and $T_{\rm i}$ as independent
variables and to determine the values $z_{\rm i}^{\rm max}=M_1/T_{\rm i}^{\rm min}$
as well as $M_1^{\rm min}$ for which the CMB constraint
$\eta_B^{\rm min}\geq\eta_{B}^{CMB}$ is satisfied. This will yield a value
$\left.M_1^{\rm min}\right|_{z_{\rm i}}$ somewhat larger than
$\left.M_1^{\rm min}\right|_{z_{\rm i}=0}$. For simplicity, we shall use the  approximation
$\left.M_1^{\rm min}\right|_{z_{\rm i}}\simeq \left.M_1^{\rm min}\right|_{z_{\rm i}=0}$
in the following.
We then define the value $z_{\rm i}^{\rm max}$, and the corresponding temperature
$T_{\rm i}^{\rm min}=M_1^{\rm min}/z_{\rm i}^{\rm max}$, by requiring that the final
asymmetry $\eta_{B}^{\rm max}$ agrees with observation within $1\s$ relative error
of the quantity $\eta_B^{CMB}/m_{\rm atm}$ which controls $M_1^{\rm min}$.

In the weak washout regime, i.e. $\mt < m_*$, one has $z_{\rm i}^{\rm max}\simeq 1$.
At temperatures smaller than $M_1$, the predicted asymmetry rapidly decreases.
Either, there is not enough time to produce neutrinos (for zero initial abundance)
or the thermal abundance is Boltzmann suppressed (for thermal initial abundance).
\begin{figure}[t]
\vspace{20mm}
\centerline{\psfig{file=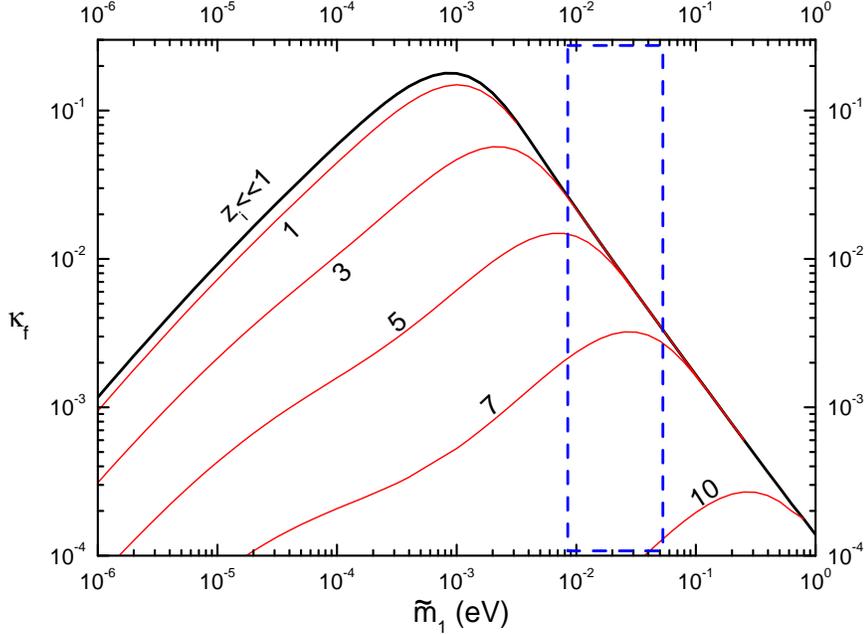,height=11cm,width=14cm}}
\vspace{-45mm}
\caption{The final efficiency factor for different values of $z_{\rm i}=M_1/T_{\rm i}$
as indicated. For $z_{\rm i}\gtrsim z_B$ there is a significant  suppression.}
\label{fig:zin}
\end{figure}
As Fig.~\ref{fig:zin} illustrates, for $\mt < m_*$ the final efficiency factors for
$z_{\rm i}=1$ and $z_{\rm i}\ll 1$ differ by only 10\%. Hence, in the weak
washout regime one has $z_{\rm i}^{\rm max}\simeq z_{\rm B} \simeq 1$.

In the strong washout regime the baryon asymmetry is predominantly produced around
$z_{\rm B}$. The value of $z_{\rm i}^{\rm max}$ is thus given by
$z_{\rm i}^{\rm max}\simeq z_{\rm B}-1.3\,\s_{\psi}$,
where $\s_{\psi}$ is the width of the Gaussian which approximates
the function $\exp{(-\psi_S^{\rm f}(z))}$ (cf.~(\ref{psis}))
peaked at $z=z_{\rm B}$ and whose integral between $z_{\rm i}^{\rm max}$ and infinity
gives the final efficiency factor minus the small error that is tolerated.
\begin{figure}
\centerline{\psfig{file=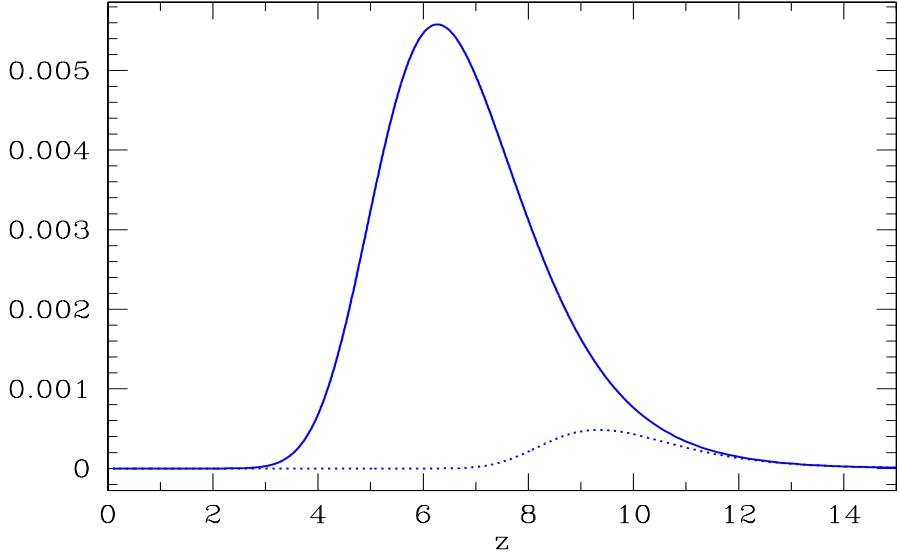,height=7.5cm,width=12cm}}
\caption{The function $\exp{(-\psi_S(z,\infty))}$ (cf.~Eq.~(\ref{psis})) for
$\mt=10^{-2}\,{\rm eV}$ (full line) and $\mt=10^{-1}\,{\rm eV}$ (dashed line).}
\label{figpsi}
\end{figure}
In Fig.~\ref{figpsi} the function $\exp{(-\psi_S^{\rm f}(z))}$ is shown for
$\mt = 10^{-2}\,{\rm eV}$ and $\mt = 10^{-1}\,{\rm eV}$, respectively. The width of
the peak is given by
$\s_{\psi} \simeq (\psi''(z_B))^{-1/2} \simeq 1.5$ for $\mt \gg m_*$.

One can easily write down an approximate expression for $z_{\rm i}^{\rm max}(K)$
which interpolates between the two regimes of weak and strong washout,
\begin{equation}\label{zinmax}
z_{\rm i}^{\rm max}(K)\simeq z_B-2\,e^{-3/K}\; .
\end{equation}
The importance of the quantity $z_B$ becomes apparent by comparing Figs.~\ref{z0} and
\ref{fig:zin}. For instance, for $\mt=m_{\rm atm}\simeq 0.05\,{\rm eV}$
one has $z_{B}\simeq 8$ and thus $z_{\rm i}^{\rm max}\simeq 6$, while
for $\mt=m_{\rm sol}\simeq 0.008\,{\rm eV}$
one has $z_{B}\simeq 6$ and thus $z_{\rm i}^{\rm max}\simeq 4$.
Clearly, for values $z_{\rm i} > z_{\rm i}^{\rm max}(K)$ the suppression of the
efficiency factor becomes significant.

From Eq.~(\ref{zinmax}) one immediately obtains a lower bound on the
initial temperature $T_{\rm i}$,
\begin{equation}
T_{\rm i}> {M_1^{\rm min}\over z_{\rm i}^{\rm max}}
\simeq {M_1^{\rm min}\over z_B-2\,e^{-3/K}}\;.
\end{equation}
The result is shown in Fig.~\ref{figbounds}.
For small $\mt \lesssim m_*$ one has
$z_{\rm i}^{\rm max}\simeq 1$
and consequently $T_{\rm i}^{\rm min}\simeq M_1^{\rm min}$.
Hence, in particular, the $3\,\sigma$ bounds (\ref{M1lbth}) and (\ref{M1lbza}) apply
also to $T_{\rm i}$.
 More interestingly, in the favored region $m_{\rm sol}\lesssim \mt\lesssim m_{\rm atm}$
 (dashed box) the $3\,\sigma$ bound (\ref{M1lbfr}) gets relaxed by a factor $4$ to $6$
 and thus
\beq
T_{\rm i}^{\rm min}=(4\times 10^9 \div 2\times 10^{10})\,{\rm GeV}
\eeq
Therefore, in the favored region of $\mt$ (dashed box),
$T_{\rm i}^{\rm min}(\mt)$ is only one order of magnitude higher than the absolute
minimum $T_{\rm i} \simeq 3\times 10^9\ {\rm GeV}$
at $\mt \simeq m_*$ (zero initial abundance) and less than
two orders of magnitude higher than the asymptotical minimum for $\mt \ll m_*$ (thermal
initial abundance). This is important in view of the `gravitino problem' which yields
an upper bound on $T_{\rm i}$ for some supersymmetric extensions of the standard model.

Comparing our results with those of \cite{gnx03}, where the additional $B-L$
asymmetry has been calculated which is produced during the reheating period at
temperatures above $T_{\rm reh}$ and below some maximal temperature $T_{\rm max}$,
we find the same amount of relaxation of the bound on $T_{\rm i}=T_{\rm reh}$.
This  indicates  that the  relaxation is a consequence of $T_B < M_1^{\rm min}$
in the case of strong washout rather than the existence of a non radiation dominated
regime above $T_{\rm reh}$.

\section{Upper bound on the light neutrino masses}

We now want to study the effect of the contribution $\D W$ to the total washout.
This term originates from the $\Delta L=2$ processes
$\f l\leftrightarrow \bar{\f} \bar{l}$ and $\f\f \leftrightarrow \bar{l} \bar{l}$
with the heavy neutrinos $N_1$, $N_2$ and $N_3$ in s- and t-channel, respectively.
$\D W$ is the only term in the kinetic equations which is not proportional to
$\mt$ but instead to the heavy neutrino mass $M_1$.

At low temperatures the washout term $\D W$ reads,
\begin{equation}\label{DWmax}
\Delta W(z) \simeq  {\o\over z^2} \left({M_1\over 10^{10}\,{\rm GeV}}\right)
\left({\mb\over {\rm eV}}\right)^2 \;,
\end{equation}
where $\mb$ is the absolute neutrino mass scale, and
the dimensionless constant $\o$ is given by
\begin{equation}
\o={9\sqrt{5}\,M_p\,10^{-8}\,{\rm GeV}^3 \over
4\pi^{9/2}\,g_l\,\sqrt{g_{\star}}\,v^4}\simeq 0.186 \; .
\end{equation}

$\D W$ is compared in Fig.~\ref{truth} of appendix~A with the total washout term
\begin{equation}
W(z)=W_0(z\, ;\mt) + \D W (z\,;M_1\mb^2) \;.
\end{equation}
As discussed in the appendix, there is a sharp transition to a low temperature regime
where $\D W$ dominates over $W_0$.
This transition occurs for a value $z_{\D} \gg 1$, which is determined by
\begin{equation}
W_0(z_{\D})=\Delta W (z_{\D}) \;.
\end{equation}
From Eqs.~(\ref{W0}) and (\ref{DWmax}) one easily obtains,
\begin{equation}
K\,z_{\D}^{9/2}\;e^{-z_{\D}}\sim \o\,
\left({M_1\over 10^{10}\,{\rm GeV}}\right)\,\left({\mb\over {\rm eV}}\right)^2 \; .
\end{equation}

In the case $z_{\D}\gtrsim z_{\rm B}$, the values of $z_{\rm B}$ and of the
efficiency factor at $z\sim z_{\rm \Delta}$ are not affected by $\D W$. Since
for $z > z_{\rm B}$ no asymmetry is produced, the total efficiency factor
is simply given by
\begin{equation}
\bar{\k}_{\rm f}(\mt,M_1\,\mb^2) =
\k_{\rm f}(\mt)\,e^{-\int_{z_{\rm B}}^{\infty}\,dz\,\Delta W(z)}\;,
\end{equation}
where the second factor describes the modification due to the presence
of $\Delta W$. Note that
$\bar{\k}_{\rm f}$ depends on $\mt$ also via $z_{\rm B}$. For
$z_{B}\gtrsim 3$ we can use the low temperature limit (\ref{DWmax})
for $\D W$, which yields
\begin{equation}
\bar{\k}_{\rm f}(\mt,\,M_1\mb^2)=\kappa_{\rm f}(\mt)
\,e^{- {\o\over z^2}\,
\left({M_1\over 10^{10}\,{\rm GeV}}\right)\,\left({\mb\over {\rm eV}}\right)^2}\;.
\end{equation}

Given the solar and atmospheric mass squared differences and a
neutrino mass pattern, i.e. $m_3^2-m_2^2 > m_2^2-m_1^2$ or
$m_3^2-m_2^2 < m_2^2-m_1^2$, the dependence of $m_3$ on $m_1$ is
fixed. In Ref.~\cite{bdp03} the absolute neutrino mass scale $\mb$
was used as variable. In the following we prefer to use instead
the lightest neutrino mass $m_1$. In the case of normal hierarchy, with
$m^{\,2}_3-m^{\,2}_2  = \D m^2_{\rm atm}$ and $m^{\,2}_2-m^{\,2}_1
= \D m^2_{\rm sol}$, one has
\begin{eqnarray} \label{numanor1}
m_3^{\,2}
&=& m_1^2 + \Delta m^2_{\rm atm} + \Delta m^2_{\rm sol}\;, \\
\label{numanor2}
m_2^{\,2} &=& m_1^2 + \Delta m^2_{\rm sol}, \\
\label{numanor3}
\mb^2  &=& 3 m_1^2 + \Delta m^2_{\rm atm} + 2\,\Delta m^2_{\rm sol}\;.
\end{eqnarray}
In the case of inverted hierarchy analogous relations hold.

Consider now the maximal baryon asymmetry (cf.~(\ref{etaB})),
\begin{equation}
\eta_{B}^{\rm max}(\widetilde{m}_1,M_1,m_1)\simeq
d\,\,\varepsilon_1^{\rm max}(\widetilde{m}_1,M_1,m_1)\,
\bar{\k}_{\rm f}(\widetilde{m}_1,M_1,m_1)\;.
\end{equation}
In the case $m_1=0$ the maximal $C\!P$ asymmetry
was depending only on $M_1$ (cf. Eq. (\ref{cpmax})).
If $m_1\geq 0$ this is suppressed by
a function $\b(\mt,m_1)\leq 1$ depending both on
$m_1$ and on $\mt$ \cite{di02,bdp03} such that
\begin{equation}
\ve_1^{\rm max}(M_1,\mt,m_1)= \ve_1^{\rm max}(M_1)\,\b(\mt,m_1)\; .
\end{equation}
The maximal value $\b=1$ is obtained in the case $m_1=0$.
The function $\b$ is conveniently factorized,
\begin{equation}
\b(\mt,m_1) = \b_{\rm max}(m_1)\,f(\mt,m_1)\; .
\end{equation}
The first factor,
\begin{equation}
\b_{\rm max}(m_1)={(m_3-m_1)\over m_{\rm atm}}={m_{\rm atm}\over (m_3+m_1)} \, ,
\end{equation}
is the maximal value of $\b$ for fixed $m_1$, which decreases $\propto 1/m_1$ for
$m_1 \gg m_{\rm atm}$. The function $f$ contains the dependence on $\mt$,
\begin{equation}\label{funf}
f(\mt,m_1)\simeq {\left(m_3-m_1\,\sqrt{1+{m^2_{\rm atm}\over
\mt^2}}\right)\over m_3-m_1}\;.
\end{equation}
This expression describes correctly the behavior of the maximal $C\!P$ asymmetry
in the limits $m_1 \rightarrow 0$ and $\mt \rightarrow \infty$. However, it has
recently been pointed out \cite{hlx03} that Eq.~(\ref{funf}) underestimates the
maximal $C\!P$ asymmetry in particular in the regime of quasi-degenerate
neutrinos\footnote{The expression (\ref{funf}) has been obtained using Eq.~(22) in
Ref.~\cite{bdp03} and assuming $x_3=0$, which is valid only in the limit
$m_1 \rightarrow 0$. Note, however, that Eq.~(\ref{funf}) approximates the maximal
$C\!P$ asymmetry within about 20\% also in the quasi-degenerate case for the relevant
values of $\mt$. For quasi-degenerate neutrinos, with $m_1 \simeq m_3 < \mt$, one easily finds that
the maximal $C\!P$ asymmetry is reached for $x_3 \simeq m_1/(2\mt)$.}.
For simplicity, we shall first calculate the neutrino mass bound
using Eq.~(\ref{funf}) and then discuss the correction.

Let us now calculate the value $M_1$ that maximizes $\eta_{B}^{\rm max}$.
In the $(\mt,M_1)$-plane this defines a trajectory $\eta_{B}^{\rm max}(\mt,m_1)$ along
which $\eta_{B}^{\rm max}$  is maximal with respect to $M_1$. The corresponding condition,
\begin{equation}
{d\ln\eta_{B}^{\rm max}\over d\,M_1}=0  \; ,
\end{equation}
yields the relation,
\begin{equation}\label{M1peak}
{\o\over z_B} \left({M_1\over 10^{10}\,{\rm GeV}}\right)
\left({\mb\over{\rm eV}}\right)^2 = 1 \;,
\end{equation}
where the quantity $z_{B}$ is a function of $\mt$. It is now easy to see that
the ratio $\eta_B^{\rm max}/\eta_B^{CMB}$, maximized with respect to $M_1$, can
be expressed in the following form,
\begin{equation}
{\eta_B^{\rm max}(\mt;m_1)\over \eta_B^{CMB}}\simeq
{\chi}\,\,\xi\,\, {z_{\rm B}(\mt)\,f(\mt,m_1)\,\k_{\rm f}(\mt)
\over (m_1+m_3)\,\mb^2 } \;,
\end{equation}
where $\chi$ is the constant
\begin{equation}
\chi={25\,d \over 6\,e\,\o}\,{{\rm eV}^4\over m_0} \simeq 1.6\,{\rm eV}^3\,\;.
\end{equation}
and $m_0=(16\,\pi/3)\,(v^2/10^{16}\,{\rm GeV})\simeq 0.051\,{\rm eV}$.
The parameter $\xi$ is the product
\begin{equation}
\xi={\xi_{\varepsilon}\,\xi^2_{\rm atm}\xi_0\over \xi_{\eta}\,\xi_{\Delta}} \; .
\end{equation}
It accounts for various factors affecting the ratio $\eta_B^{\rm max}/\eta_B^{CMB}$:
(1) the maximal $C\!P$ asymmetry,
$\xi_{\ve}$; (2) the atmospheric neutrino mass scale,
$\xi_{\rm atm}=m_{\rm atm}/(0.05\,{\rm eV})$;
(3) the observed baryon asymmetry, $\xi_{\eta}=\eta_{B}^{CMB}/(6\times 10^{-10})$;
(4) the variation $\xi_{\D}$ of the strength of the $\D L=2$ washout term, and (5) the
variation $\xi_{0}$ of the efficiency factor at small $M_1\mb^2$.
This parametrization of the maximal asymmetry is
useful to study the dependence of the neutrino mass bound on the various parameters
involved.

In order to determine the absolute maximum of the asymmetry $\eta_B^{\rm max}$
we also have to find the extremum with respect to $\mt$ and, finally, the maximum
with respect to $m_1$ or, equivalently, the absolute neutrino mass scale.
Comparison with the observed asymmetry $\eta^{CMB}_B$ then yields the leptogenesis
neutrino mass bound. Anticipating again that the maximum falls in the region
of large $\mt$, we can use the analytical expression (\ref{Skfstrong}) for $\k_{\rm f}$
in the strong washout region. Since $j(z_B \gg 1) \simeq 1$, one has
\begin{equation}\label{zBk}
z_{\rm B}(\mt)\,\k_{\rm f}(\mt) \simeq {2\over K}={2\,m_*\over \mt} \;.
\end{equation}
Further, for large $\mt$ the function $f(\mt,m_1)$ can be approximated by
\begin{equation}\label{fapp}
f(\mt,m_1) \simeq 1-{1\over 2}{(m_3+m_1)\,m_1\over\mt^2} \;.
\end{equation}
With this simplified expression
it is easy to see that the peak is reached for
\begin{equation}\label{mt}
\mt^{\rm peak} \simeq \sqrt{{3\over 2}\,m_1\,(m_1+m_3)} \; ,
\end{equation}
corresponding to $f(\mt^{\rm peak},m_1)=2/3$.
The peak value of the asymmetry is given by
\begin{equation}
{\eta_{B}^{\rm peak}(m_1)\over \eta_B^{CMB}} \simeq {2^{5/2}\over 3^{3/2}}\,
{\chi\,m_*\,\xi \over (m_1+m_3)^{3/2}\,m_1^{1/2}\,\mb^2} \;.
\end{equation}

Anticipating $(m_1^{\rm peak})^2\gg m^2_{\rm atm}$, one has to zeroth order in
$(m_{\rm atm}/m_1)^2$,
\begin{equation}\label{m}
m_3^0 = {\mb^0\over\sqrt{3}}\simeq m_1\;.
\end{equation}
Imposing now the CMB constraint $\eta_B\geq\eta_B^{CMB}$ we find the leptogenesis bound
on the absolute neutrino mass scale (cf.~\cite{bdp022,bdp03}),
\begin{equation}
m_i \leq m^0_{\rm peak}=\sigma\,\xi^{1/4}\,{\rm eV}\;,
\end{equation}
with
\begin{equation}
\sigma =10^6 \,
\left({10\,g_{\rm rec}\,a_{\rm sph}\,\pi^6\over 3^{9/2}\,e}\right)^{1/4}\,
\left({v^2\over M_{\rm PL}\,{\rm GeV}}\right)^{1/2}\simeq 0.121 \;.
\end{equation}
In this last equation we used the fact that in a standard thermal
history the dilution factor, contained in $d$, is given by
$N_{\gamma}^{\rm rec}/N_{\gamma}^{\star}=g_*/g_{\rm rec}$
with $g_{\rm rec}=43/11$ the number of the (entropy) degrees of freedom
at recombination. Combining Eqs.~(\ref{m}) and (\ref{mt}) one finds
\begin{equation}\label{max}
\widetilde{m}_1^{\rm peak}\simeq \mb^0_{\rm peak} \simeq 0.17\,{\rm eV} \;,
\end{equation}
which is consistent with the approximation of strong washout used in Eq.~(\ref{zBk}).
From Fig.~\ref{z0} one then reads off $z_B^{\rm peak}\simeq 10$.
Together with (\ref{M1peak}) this yields for the peak value of $M_1$,
\begin{equation}
M_1^{\rm peak} =
{z_{\rm B}(\mt^{\rm peak})\over \o\,(\mb^0_{\rm peak})^2}
\simeq 2\times 10^{13}\,\xi^{-1/2}\,{\rm GeV}\;.
\end{equation}

It is straightforward to go beyond the zeroth order in $(m_{\rm atm}/m_1)^2$. In the
case of normal hierarchy the lightest neutrino mass bound is given by
\begin{equation}
m^{\rm peak}_{\rm 1, nor}\simeq m_{\rm peak}^0\,
\left(1-{17\over 96}\,{m^2_{\rm atm}\over (m_{\rm peak}^0)^2}\right) \;,
\end{equation}
whereas in the case of inverted hierarchy one has
\begin{equation}
m^{\rm peak}_{\rm 1, inv}\simeq
m_{\rm peak}^0\,\left(1-{25\over 96}\,{m^2_{\rm atm}\over
(m_{\rm peak}^0)^2} \right) \;,
\end{equation}
which yields
$m^{\rm peak}_{\rm 1, nor}\simeq m^{\rm peak}_{\rm 1, inv}
\simeq m_{\rm peak}^0-0.005\,{\rm eV}$.
In order to obtain numerical results for the upper bounds on the light neutrino masses
one has to specify the baryon asymmetry and the neutrino mass squared differences.
For $\eta_B^{CMB}$ we use the WMAP plus SDSS result (\ref{etanew}), while the value for
$m_{\rm atm}$ is given by the Eq.~(\ref{matm}).
Since $\mb_{\rm peak}\propto (m^2_{\rm atm}/\eta_{CMB})^{1/4}$, the experimental
error on $m_{\rm peak}^0$ is about $5\%$. Setting all other parameters $\xi_i=1$,
one finds for the central value $\xi=\xi^2_{\rm atm}/\xi_{\eta}\simeq 0.95$.
We then obtain
\begin{equation}
m_1^{\rm peak}=(0.115 \pm 0.005)\,{\rm eV}\; .
\end{equation}
The corresponding $3\sigma$ upper bounds on the neutrino masses are for normal hierarchy,
\begin{equation}\label{bnor}
m_1,m_2 < 0.13\,{\rm eV}\; ,\;\;\; m_3 < 0.14\,{\rm eV}\; ,
\end{equation}
and correspondingly in the case of inverted hierarchy,
\begin{equation}\label{binv}
m_1< 0.13\,{\rm eV}\; ,\;\;\; m_2,m_3 < 0.14\,{\rm eV}\; .
\end{equation}
These analytical bounds are consistent with the numerical results
obtained in \cite{bdp03}, if one accounts for the different parameters,
$m_{\rm atm}=0.05\,{\rm eV}$ and $\eta_{CMB}=3.6\times 10^{-10}$ ($\xi \simeq 1.7$),
and the over-estimate of the washout term $W_0$ by 50\%.

The bound on the maximal $C\!P$ asymmetry derived in \cite{hlx03} corresponds
in the relevant range of large $m_1$ and $\mt$ to the function (cf.~(\ref{funf})),
\begin{equation}\label{funfcor}
f(\mt,m_1)\simeq \sqrt{1-{m_1^2\over \widetilde{m}_1^2}}\;.
\end{equation}
Repeating the above analysis one finds that the peak of the asymmetry is shifted
to $\mt^{\rm peak}=\sqrt{2}m_1$, with $f^{\rm peak}\simeq 1/\sqrt{2}$.
From Eqs.~(\ref{zBk}), (\ref{m}) and (\ref{max}) one then concludes that the neutrino
mass bound is relaxed by the factor $(3^{3/2}/4)^{1/4}\simeq 1.07$, i.e. 7\%,
corresponding to an increase of the neutrino mass bound by $0.01\,{\rm eV}$ .

An important correction arises from the dependence of the neutrino masses
on the renormalization scale $\m$.
The only low energy quantity upon which  $m_1^{\rm peak}$ depends
is the atmospheric neutrino mass scale $m_{\rm atm}$. Hence,
there are two competing effects: the first  one is the running of $m_{\rm atm}$ from
the Fermi scale $\m=m_Z$ to the high scale $\m \sim M_1$ ($\sim 10^{13}\,{\rm GeV}$),
the second one is the running of $m_1^{\rm peak}$ from $\m \sim M_1$ down to
$\m = m_Z$. In the standard model the light neutrino masses scale uniformly under
the renormalization group. Since $m_{\rm peak}\propto \sqrt{m_{\rm atm}}$,
the first effect then gives a correction that is only half of the second one.
Renormalization group effects make the bound more restrictive \cite{akx03}.
In order to have an upper bound, we have to choose those values
of the parameters that produce the smallest effect.
This corresponds to choosing the lowest Higgs mass  compatible with
positive Higgs self-coupling at $\sim 10^{13}\,{\rm GeV}$, which is about
$150\,{\rm GeV}$. The atmospheric neutrino mass scale is then increased
by about $40\%$ \cite{akx03} and the bound gets $20\%$ weaker at $\m \sim M_1$,
but $20\%$ more restrictive at $\m = m_Z$. Combining the effect of radiative corrections
and the larger $C\!P$ asymmetry (Eq.~(\ref{funfcor})), we finally obtain from
Eq.~(\ref{bnor}) and (\ref{binv}) at $3\sigma$,
\begin{equation}
m_i < 0.12 \, {\rm eV} \;,
\end{equation}
which, thanks to cancellations among different corrections, agrees with
the bound obtained in \cite{bdp03}.

It is important to realize, however, that there are corrections of the same order as
those discussed above which cannot be treated within the present framework.
It is usually assumed that in leptogenesis first a lepton asymmetry is produced, which is
then partially transformed into a baryon asymmetry by sphaleron processes. However, this
picture is incorrect \cite{bp01}. The duration of leptogenesis is about two orders of
magnitude larger than the inverse Hubble parameter when it starts. Since many processes
in the plasma, in particular the sphaleron processes, are much faster, the generated
asymmetry is `instantaneously' distributed among quarks and leptons. Hence, the chemical
potentials of quarks and leptons are changed already during the process of
leptogenesis. A complete analysis has to take into account how the contributing `spectator
processes', which are in thermal equilibrium, change with decreasing temperature
(cf.,~e.g.,~\cite{bp00}). In \cite{bp01} is has been estimated that spectator processes
reduce the generated baryon asymmetry by about a factor of two. Hence, there is presently
a theoretical uncertainty of at least $0.02\,{\rm eV}$ on the neutrino mass bound.

Finally, it has to be kept in mind that our whole analysis is based on the simplest
version of the seesaw mechanism with hierarchical heavy Majorana neutrinos.
The leptogenesis neutrino mass bound can be relaxed if the heavy Majorana neutrinos are,
at least partially, quasi-degenerate in mass. In this case the $C\!P$ asymmetry
can be much larger \cite{fps95,crv96} than the upper bound used in the above discussion.
This possibility has to be discussed in the context of realistic models of neutrino
masses. Further, if Higgs triplets contribute significantly to neutrino masses the
connection between baryon asymmetry and neutrino masses disappears entirely.
Different relations between neutrino masses and baryon asymmetry are also obtained
in non-thermal leptogenesis \cite{ls91,hmy02}.

\section{Towards the theory of thermal leptogenesis}

The goal of leptogenesis is the prediction of the baryon asymmetry, given neutrino masses
and mixings. The consistency of present calculations with observations is impressive, but
so far it is not possible to quote a rigorous theoretical error on the predicted
asymmetry, which is a necessary requirement for a `theory of leptogenesis'.

The generation of a baryon asymmetry is a non-equilibrium process which is generally
treated by means of Boltzmann equations. This procedure has a basic conceptual problem:
the Boltzmann equations are classical equations for the time evolution of phase space
distribution functions; the involved collision terms, however, are zero temperature
S-matrix elements which involve quantum interferences in a crucial manner. Clearly, a
full quantum mechanical treatment is necessary to understand the range of validity of
the Boltzmann equations and to determine the size of corrections.

A first step in this direction has been made in \cite{bf00} where a perturbative
solution of the exact Kadanoff-Baym equations has been constructed. To zeroth order,
for non-relativistic heavy neutrinos, the non-equilibrium Green functions have
been obtained in terms of distribution functions satisfying Boltzmann equations.
Note that in the favoured strong washout regime the decaying heavy neutrinos are
indeed non-relativistic. It is instructive to recall the various corrections.
There are off-shell contributions, `memory effects' related to the derivative expansion
of the Wigner transforms, relativistic corrections and higher-order loop corrections.
All these correction terms are known explicitly, but their size during the process of
baryogenesis and, in particular, their effect on the final baryon asymmetry have not
yet been worked out.

Recently, thermal corrections have been studied \cite{gnx03}. They correspond to loop
corrections involving gauge bosons and the top quark.
At large temperatures, $T > M_1$, the processes in the plasma and the
$C\!P$ asymmetries change significantly if thermal masses are treated as kinematical
masses in the evaluation of scattering matrix elements \cite{gnx03}. On the
contrary, thermal corrections are small if they are only included as propagator
effects \cite{crx98}. To clarify this issue is of general importance for the treatment
of non-equilibrium processes at high temperatures.

The effect of all these corrections on the final baryon asymmetry depends crucially on
the value of the neutrino masses. Large thermal corrections would modify the asymmetry at
temperatures above $M_1$. This affects the final baryon asymmetry only in the case of
small washout, i.e. $\mt < m_*$. In the strong washout regime, $\mt > m_*$, which appears
to be favored by the current evidence for neutrino masses, the baryon asymmetry is
generated at a temperature $T_B < M_1$. In this case thermal corrections are small.
Correspondingly, the recently obtained bounds on light and heavy neutrino masses
\cite{bdp03,gnx03,hlx03} are all very similar.

The final value of the baryon asymmetry is significantly affected by `spectator
processes' \cite{bp01} which cannot be treated based on the simple Boltzmann
equations discussed in this paper. It has been estimated that this effect changes
the baryon asymmetry by a factor of about two, leading to a theoretical uncertainty
of the leptogenesis neutrino mass bound of about $0.02\,{\rm eV}$.
Clearly, to obtain a more accurate prediction for
the baryon asymmetry requires a considerable increase in the complexity of the calculation.

An important step towards the theory of leptogenesis would be a systematic evaluation of
all corrections to the simple Boltzmann equations in the `easy regime' of strong washout
where $\mt > m_*$. One could then see where this approach breaks down as $\mt$ decreases
and $T_B$ approaches $M_1$. On the experimental side, information on the absolute neutrino
mass scale $\mb$, and therefore on $\mt > m_1$, is of crucial
importance. Maybe, we are lucky, and nature has chosen neutrino masses in the strong
washout regime where leptogenesis works best.\\

\vspace{-5mm}
\noindent
{\bf Acknowledgments}\\
We would like to thank L.~Covi, K.~Hamaguchi, J.~Pati and M.~Ratz for
helpful discussions, and G. Kane and the Michigan Center for Theoretical
Physics (MCTP) for hospitality during the {\it Baryogenesis Workshop}.
P.D.B. thanks the DESY and CERN theory groups for their kind hospitality.
The work of P.D.B. has been supported by the EU Fifth Framework network ``Supersymmetry
and the Early Universe" (HPRN-CT-2000-00152).

\newpage

\section*{Appendix A}

A crucial and delicate point in setting up the Boltzmann equations for
leptogenesis is the subtraction of the real intermediate state
contribution (RIS) from the $2\rightarrow 2$ scattering processes
\cite{kw80}. Without this subtraction, decays and inverse decays lead
to the generation of a lepton asymmetry in thermal equilibrium, in
contradiction with general theorems.

In order to explicitly split the $2\rightarrow 2$
scattering processes into RIS and remainder one has to calculate the
$\D L=2$ processes, including one-loop self-energy and vertex
corrections in the resonance region. This calculation has been carried
out in \cite{bp98} where the relevant results are given in Eqs.~(68) -
(85).

Let us consider for simplicity the case $s, M_1^2 \ll M_2^2, M_3^2$
($s$, $t$ and $u$ are the usual Mandelstam variables). For our
purposes it is sufficient to study the averaged matrix element
squared,
\bea
|{\cal M}(\bar{l}\bar{\f}\rightarrow l\f)|^2_{av} =
\int_{-s}^0 du |{\cal M}(\bar{l}\bar{\f}\rightarrow l\f)|^2 \,,
\eea
where the integral over $u$ corresponds to the integral over the final state
lepton angle, i.e. a partial phase space integration.

In order to study the resonance region the diagonal part of the
self-energy was re-summed in \cite{bp98} whereas the off-diagonal part
was treated as perturbation in the Yukawa couplings $h$.  In the
$2\rightarrow 2$ scattering amplitude the free propagator is then
replaced by a Breit-Wigner propagator,
\bea
{1\over s - M_1^2 + i M_1 \G_1} =
{1\over M_1^2}\left({1\over D_1(x)} -i {1\over R_1(x)}\right)\;,
\eea
where
\bea
{1\over D_1(x)} = {x-1\over (x-1)^2 + c^2}\;, \quad
{1\over R_1(x)} = {c\over (x-1)^2 + c^2}\;,
\eea
and
\bea\label{unit}
{1\over D_1(x)^2} + {1\over R_1(x)^2} = {1\over c}{1\over R_1(x)}\; ,
\eea
with
\bea
x = {s\over M_1^2}\;, \quad c = {\G_1\over M_1} = {1\over 8\p}K_{11}\;, \quad
K_{ij} = (h^\dg h)_{ij}\; ;
\eea
here $h_{kj}$ is the Yukawa coupling of $N_j$ to $l_k\f$.

The averaged matrix elements are then given by the following expression \cite{bp98},
\bea
|{\cal M}(\bar{l}\bar{\f}\rightarrow l\f)|^2_{av} &=&
2 s^2 \sum_{ij}\left(A_{ij} - B_{ij} - C_{ij} -
4s \sum_k\left(D_{ijk} + \overline{E}_{ijk}\right)\right)\;, \\
|{\cal M}(l\f\rightarrow \bar{l}\bar{\f})|^2_{av} &=&
2 s^2 \sum_{ij}\left(A_{ij} + B_{ij} + C_{ij} +
4s \sum_k\left(D_{ijk} + \overline{E}_{ijk}\right)\right)\;,
\eea
where we have only shown terms contributing to the subtraction of the RIS
part as well as the leading order off-shell part, e.g.\ $C\!P$ conserving
one-loop corrections have not been included.
$A_{ij}$ and $B_{ij}$ represent the various $N_i$-$N_j$ s-channel interference terms.
Up to terms ${\cal O}(h^6)$ they are ($i,j=2,3$)
\bea
A_{11} &=& K_{11}^2 {1\over M_1^2}\left({1\over D_1^2} + {1\over R_1^2}\right)\;, \\
A_{1i} + A_{i1} &=& -2\ {\rm Re}\{K^2_{1i}\}{1\over M_1 M_i} {1\over D_1}\;, \\
A_{ij} &=& {\rm Re}\{K^2_{ij}\}{1\over M_i M_j}\;, \\
B_{11} &=& B_{ij} + B_{ji} = 0 \;,\\
B_{1i} + B_{i1} &=& 2\ {\rm Im}\{K^2_{1i}\}{1\over M_1 M_i} {1\over R_1}\;;
\eea
the $N_1$-$N_1$ s-channel terms with self-energy and vertex corrections, respectively,
read ($k=2,3$)
\bea
C_{1k} &=& - 2x\ {\rm Im}\{K_{1k}^2\} {c\over M_1 M_k}
           \left({1\over D_1^2} + {1\over R_1^2}\right) \;,\\
D_{11k} &=&{1\over 2}\ {\rm Im}\{K_{1k}^2\}
           {c\over \sqrt{x}M_1^4} f\left({M_k^2\over s}\right)
           \left({1\over D_1^2} + {1\over R_1^2}\right)\;,
\eea
where
\bea
f(y) &=& \sqrt{y}\left(1-(1+y)\ln{\left({1+y\over y}\right)}\right) \NO\\
     &=& - {1\over 2\sqrt{y}} + {\cal O}\left({1\over y}\right)\;;
\eea
finally, the s-u-channel interference term is
\bea
\overline{E}_{11k} = - D_{11k}\;.
\eea

From these equations one reads off
\bea
  D_{11k} + \overline{E}_{11k} = 0 \; ,
\eea
and, for $s=M_1^2$,
\begin{equation}
  B_{1i} + B_{i1} + C_{1i} = 0\;.
\end{equation}
Hence, the $C\!P$ asymmetry of the full $2\rightarrow 2$ cross section
vanishes to ${\cal O}(h^4)$. The `pole terms', corresponding to $N_1$-$N_1$ s-channel
contributions, are cancelled by on-/off-shell s-channel interferences (self-energy)
and s-channel/u-channel interference (vertex correction). Off-shell, the corresponding
cancellations take place to ${\cal O}(h^6)$ \cite{rcv98}.

As an unstable particle the heavy neutrino $N_1$ is defined as pole in the
$2\rightarrow 2$ scattering amplitude
\bea
{\cal M}(\bar{l}\bar{\f}\rightarrow l\f) \simeq
\langle l\f|N_1\rangle {i\over s - M_1^2 + iM_1\G_1} \langle N_1|\bar{l}\bar{\f}\rangle \;.
\eea
The residue yields the decay amplitude and, in particular, the $C\!P$ asymmetry.
The RIS term can then be identified as the squared matrix element in the zero-width limit,
\bea
|{\cal M}(\bar{l}\bar{\f}\rightarrow l\f)|^2_{RIS} &=&
\lim_{\G_1 \rightarrow 0}\, 2s^2\left (A_{11}
- \sum_k \left(C_{1k} + 4sD_{11k}\right)\right) \NO\\
&=& 16 \p^2 K_{11} M_1^2 \left(1 + 2\ve_1^M + 2\ve_1^V\right) \d(x-1) \;,
\eea
where $\ve_1^M$ and  $\ve_1^V$ are the familiar $C\!P$ asymmetries due to mixing and
vertex correction, respectively ($M_1 \ll M_2, M_3$),
\bea
\ve_1^M &=& {1\over 8\p} \sum_k {{\rm Im}\{K_{1k}^2\}\over K_{11}}{M_1\over M_k}\;,\\
\ve_1^V &=& - {1\over 8\p} \sum_k {{\rm Im}\{K_{1k}^2\}\over K_{11}}
              f\left({M_k^2\over M_1^2}\right) \;.
\eea
In \cite{kw80} it has been shown that the subtraction of the RIS term, which corresponds
to the replacement $(1/D_1^2 + 1/R_1^2) \rightarrow (1/D_1^2 + 1/R_1^2) - (\p/c) \d(x-1)$,
leads to Boltzmann equations with the expected properties, which have the equilibrium
solution $N_1 = N_1^{\rm eq}$, $N_{B-L}=0$.

It is now straightforward to write down the subtracted matrix element squared. Keeping
only terms ${\cal O}(h^4)$, where the zero-width limit can be taken for $B_{1i}$ and
$\overline{E}_{11k}$, one obtains the simple expressions ($i,j=2,3$),
\bea
|{\cal M}(\bar{l}\bar{\f}\rightarrow l\f)|^2_{sub} &=&
|{\cal M}_{\D L=2}|^2_+ + |{\cal M}_{\D L=2}|^2_- \;,\\[1ex]
|{\cal M}(l\f\rightarrow \bar{l}\bar{\f})|^2_{sub} &=&
|{\cal M}_{\D L=2}|^2_+ - |{\cal M}_{\D L=2}|^2_- \;,
\eea
with
\bea
    \left|{\cal M}_{\Delta L=2}\right|^2_+
&=& 2s^2\left\{{K_{11}^2\over M_1^2}\left[
    {1\over D_1^2}+{1\over R_1^2}-{\pi\over c}\delta(x-1)\nonumber\right.\right.\\[1ex]
&&\qquad\qquad \left.+\;
{2\over x}-{2\over x^2}\left(1+{x+1\over D_1}\right)
    {\rm ln}(x+1)+{2\over xD_1}\right]
    \nonumber\\[1ex]
&& -6\;\sum\limits_{i}{\rm Re}\left(K_{1i}^2\right){1\over M_1M_i}\left[
   {1\over x}+{1\over 2D_1}-{(x+1)\over x^2}
   {\rm ln}(x+1)\right]\nonumber\\[1ex]
&& \left.+\;3\sum\limits_{i,j}{\rm Re}\left(K_{ij}^2\right){1\over M_iM_j}\right\}
   \; , \label{correct1}\\[1ex]
|{\cal M}_{\D L=2}|^2_- &=& {1\over 2}\left(
|{\cal M}(\bar{l}\bar{\f}\rightarrow l\f)|^2_{av} -
|{\cal M}(l\f\rightarrow \bar{l}\bar{\f})|^2_{av}\right) \NO\\[1ex]
&=&-32\pi^2 K_{11} M_1^2(\ve_1^M + \ve_1^V) \d(x-1)\;. \label{correct2}
\eea
Note that the subtracted squared matrix element, contrary to the unsubtracted one,
violates $C\!P$. To leading order in the coupling this part contributes only
on-shell, and it is ${\cal O}(h^2)$ suppressed with respect to the leading Born term.
Away from the pole, for $s \ll M_1^2$, one has
\bea
|{\cal M}(\bar{l}\bar{\f}\rightarrow l\f)|^2_{sub} &=&
|{\cal M}(l\f\rightarrow \bar{l}\bar{\f})|^2_{sub} \NO\\
&=& 6 s^2 \sum_{ij}{\rm Re}\{K^2_{ij}\}{1\over M_i M_j} \NO\\
&=& {6 s^2\over v^4} {\rm tr}(m_\n^\dg m_\n) \;,
    \label{lowenergy1}
\eea
which is the crucial term leading to the upper bound on the neutrino masses
\cite{bdp02}.

We also have to take into account the $\D L=2$ process $ll\rightarrow \bar{\f}\bar{\f}$.
The corresponding matrix elements reads
\bea
    \left|{\cal M}_{\Delta L=2,t}\right|^2
&=& 2s^2\left\{ {K_{11}^2\over M_1^2}\left[{2\over x+1}+
    {2\over x(x+2)}\ln(x+1)\right]\right.\nonumber\\[1ex]
&&  \left.+6\sum\limits_i{\rm Re}\left(K_{1i}^2\right){1\over M_1 M_i}{1\over x}
    \ln(x+1)
    +3\sum\limits_{i,j}{\rm Re}\left(K_{ij}^2\right){1\over M_i M_j}\right\}.
\eea
For small center of mass energies one again obtains
\begin{equation}
  \left|{\cal M}_{\Delta L=2,t}\right|^2={6s^2\over v^4}{\rm tr}(m_\n^\dg m_\n) \;.
  \label{lowenergy2}
\end{equation}

For the derivation of the upper bound on the light neutrino masses one needs the maximal
$C\!P$ asymmetry for given $M_1$, $\mt$ and $\mb$. In this case also the complete
$\D L = 2$ matrix element depends just on these three variables. This is easily seen in the
flavor basis where the Yukawa matrix $\th$ connects light and heavy neutrino mass
eigenstates. The matrix
\bea\label{ortho}
\O_{ij} = {v\over \sqrt{m_iM_j}}\th_{ij}\;,
\eea
is then orthogonal, $\O \O^T = \O^T \O = I$ \cite{ci01}, which implies
\bea
\sum_k \th_{ik}\th_{jk}{1\over M_k} = {\sqrt{m_i m_j} \over v^2} \d_{ij} \;.
\eea
Using ${\rm Re}\{K^2_{1i}\} = {\rm Re}\{K^2_{i1}\}$, this implies
for the interference term appearing in Eq.~(\ref{correct1}),
\bea
\sum_{i\neq 1}{\rm Re}\{K^2_{1i}\}{1\over M_i} =
-{K_{11}^2\over M_1} + \sum_{j=1}^3 {m_j\over v^2}{\rm Re}\{\th^2_{j1}\} \;.
\eea
%As shown in \cite{bdp03}, the maximal $C\!P$ asymmetry is reached for
The conditions
\bea\label{cond}
{\rm Re}\{\th^2_{21}\} = {\rm Re}\{\th^2_{31}\} = 0\;, \quad
{\rm Re}\{\th^2_{11}\} = {m_1 M_1\over v^2}\; ,
\eea
yield a good approximation for the maximal $C\!P$ asymmetry \cite{bdp03}. The difference
to the maximal $C\!P$ asymmetry \cite{hlx03} can then be treated as a perturbation, as
discussed in sect.~4. Eq.~(\ref{cond}) then implies for the interference term
\bea
\sum_{i\neq 1}{\rm Re}\{K^2_{1i}\}{1\over M_i} = -
{M_1\over v^4}\left(\mt^2 - m_1^2\right)\;.
\label{int}
\eea
Inserting this expression in Eq.~(\ref{correct1}) one obtains for the $\D L=2$ matrix
element in the case of maximal $C\!P$ asymmetry,
\bea
\left|{\cal M}_{\Delta L=2}\right|^2_+
&=& {2s^2\over v^4}\left\{\widetilde{m}_1^2\left[
    {1\over D_1^2}+{1\over R_1^2}-{\pi\over c}\delta(x-1)+
    {2\over x}\left(1+{1\over D_1}\right)-3 \right.\right.\\[1ex]
&&  \qquad\qquad\left.-{2\over x^2}\left(1+{x+1\over D_1}\right)
    \ln(x+1)\right]\nonumber\\[1ex]
&&  \left.+6\left(\widetilde{m}_1^2-m_1^2\right)\left[{x+1\over x}
    +{1\over 2D_1}-{(x+1)\over x^2}
    \ln(x+1)\right]+3\overline{m}^2\right\}\;. \nonumber
\eea
For the process $ll\rightarrow \bar{\f}\bar{\f}$ one obtains in the case of
maximal $C\!P$ asymmetry,
\bea
    \left|{\cal M}_{\Delta L=2,t}\right|^2
&=& {2s^2\over v^4}\left\{\widetilde{m}_1^2\left[
    {2\over x+1}-3+{2\over x(x+2)}\ln(x+1)
    \right]\right.\nonumber\\[1ex]
&&  \left.+6\left(\widetilde{m}_1^2-m_1^2\right)\left[1-
    {1\over x}\ln(x+1)\right]+3\overline{m}^2\right\}\; .
\eea

For small energies, $s\ll M_1^2$, these matrix elements again reduce to
\begin{equation}
  \left|{\cal M}_{\Delta L=2}\right|^2_+=\left|{\cal M}_{\Delta L=2,t}\right|^2=
  {6s^2\over v^4}\overline{m}^2\;,
  \label{lowenergy}
\end{equation}
whereas for intermediate energies $M_1^2\ll s\ll M_{2,3}^2$ one finds
\begin{equation}
  \left|{\cal M}_{\Delta L=2}\right|^2_+=\left|{\cal M}_{\Delta L=2,t}\right|^2
  ={6s^2\over v^4}\left(\mt^2+\mb^2-2m_1^2\right)\;.
  \label{intermediate}
\end{equation}

Following \cite{kw80,hkx82}, it has been standard practice \cite{lut92}-\cite{pu03}
to determine $|{\cal M}|^2_{sub}$ by computing the
Born diagrams for the $2\rightarrow 2$ process with Breit-Wigner propagator and
dropping $1/R_1$, the imaginary part of the propagator, since in the zero width limit
\bea
{1\over s - M_1^2 + i M_1 \G_1} &=&
{1\over M_1^2}\left({1\over D_1(x)} -i {1\over R_1(x)}\right) \NO\\
&\stackrel{\G_1 \rightarrow 0}{\longrightarrow}& - i \p \d(s-M_1^2)\;.
\eea
Recently, it has been pointed out that this procedure is not correct
\cite{gnx03}. In a toy model, the same conclusion has been reached in \cite{saw03}.
Indeed, the described procedure leads to a subtracted squared matrix
element which contains terms $\propto 1/D_1^2$ \cite{hkx82}, implying
$|{\cal M}|^2_{sub} = {\cal O}(h^2)$ on-shell, in contradiction with
Eq.~(\ref{correct1}). The zero-width limits of the squared amplitude
and the squared imaginary part are different.
This was overlooked in the past, leading to an overestimate of the washout rate
due to inverse decays by 50\% \cite{gnx03}. The RIS term has to be subtracted from the
full $2\rightarrow 2$ cross section, not just from the Born cross section, in order to
obtain the crucial $C\!P$ violating contribution proportional to $\ve_1$\footnote{
Note that in Ref.~\cite{gnx03}
$\g_{N_s}^{\rm on-shell}(LH\rightarrow \bar{L}\bar{H})$ is $C\!P$ violating, using
a given $C\!P$ asymmetry $\e_{N_1}$ not determined by the $2\rightarrow 2$ processes.
It is different from the on-shell part of $\g_{N_s}(LH\rightarrow \bar{L}\bar{H})$
which, as a tree-level rate, conserves $C\!P$. As discussed above, the correct
subtraction term is obtained from the full reaction rate including vertex and self-energy
corrections to ${\cal O}(h^4)$ by separating the on-shell part from the interference terms.
This procedure automatically yields the correct $C\!P$ asymmetry $\e_{N_1}$.}.

Let us now consider the Boltzmann equation for the density of lepton doublets,
assuming kinetic equilibrium,
\begin{eqnarray}
{dn_l\over dt} + 3 H n_l
& = & {n_{N_1} \over n^{\rm eq}_{N_1}}\gamma^{\rm eq}(N_1\to l\phi)
               - {n_l \over n^{\rm eq}_l}\gamma^{\rm eq}(l\phi\to N_1)
               \\[1ex]
          && +\; {n_{\bar{l}} \over n^{\rm eq}_l}
                 \gamma^{\rm eq}_{\rm sub}(\bar{l}\bar{\phi}\to l\phi)
               - {n_l \over n^{\rm eq}_l}
                 \gamma^{\rm eq}_{\rm sub}(l\phi\to\bar{l}\bar{\phi})
               \nonumber\\[1ex]
          && +\; \gamma^{\rm eq}(\bar{\phi}\bar{\phi}\to ll)
               - \left({n_l \over n^{\rm eq}_l}\right)^2
                 \gamma^{\rm eq}(ll\to\bar{\phi}\bar{\phi})
               \nonumber\;.
\end{eqnarray}
Here, $\g^{\rm eq}$ are the usual reaction densities in thermal equilibrium and we
have assumed that the Higgs doublets $\phi$ are in thermal equilibrium, neglecting
their chemical potential.
%From Eqs.~() and (\ref{correct}) one reads off
The $C\!P$ asymmetry $\varepsilon_1$ is defined in such a way that
\begin{eqnarray}
  \gamma^{\rm eq}(N_1\to l\phi) =
  \gamma^{\rm eq}(\bar{l}\bar{\phi}\to N_1) &=&
  {1+\varepsilon_1\over 2}\gamma_{N_1}\;,\\[1ex]
  \gamma^{\rm eq}(N_1\to \bar{l}\bar{\phi}) =
  \gamma^{\rm eq}(l\phi\to N_1) &=&
  {1-\varepsilon_1\over 2}\gamma_{N_1}\;.
\end{eqnarray}
Further, for the $2\to2$ processes we have
\begin{eqnarray}
  \gamma^{\rm eq}_{\rm sub}(\bar{l}\bar{\phi}\to l\phi)
  &=& \gamma^{\rm eq}_{\Delta L=2,+}
          -{1\over2}\varepsilon_1\gamma_{N_1}\;,\\[1ex]
  \gamma^{\rm eq}_{\rm sub}(l\phi\to\bar{l}\bar{\phi}) &=& \gamma^{\rm eq}_{\Delta L=2,+}
          +{1\over2}\varepsilon_1\gamma_{N_1}\;,\\[1ex]
  \gamma^{\rm eq}(\bar{\phi}\bar{\phi}\leftrightarrow ll)&=&
  \gamma^{\rm eq}(\bar{l}\bar{l}\leftrightarrow\phi\phi) = \gamma^{\rm eq}_{\Delta L=2,t}\;.
\end{eqnarray}
Introducing a lepton, or $B-L$ asymmetry,
\bea
n_l = n_l^{\rm eq} - {1\over 2} n_{B-L}\;, \quad
n_{\bar{l}} = n_l^{\rm eq} + {1\over 2} n_{B-L}\;,
\eea
assuming $n_{B-L} = n_{\bar{l}} - n_l = {\cal O}(\ve_1)$, and keeping only terms
${\cal O}(\ve_1)$, one obtains the kinetic equation for the $B-L$ asymmetry
\bea\label{boltzm}
{dn_{B-L}\over dt} + 3 H n_{B-L}
= -\ve_1\left(\frac{n_{N_1}}{n_{N_1}^{\rm eq}}-1\right)\ \g_{N_1}
- {n_{B-L} \over n_l^{\rm eq}}\ \left({1\over 2} \g_{N_1} + \g_{\D L=2} \right)\;,
\eea
where
\begin{equation}
  \gamma_{\Delta L=2} = 2\,\gamma^{\rm eq}_{\Delta L=2,+}
                      + 2\,\gamma^{\rm eq}_{\Delta L=2,t}\;.
\end{equation}
The $C\!P$ violating part of $\g_{\rm sub}^{\rm eq}$ yields the term $+\ve_1\g_{N_1}$
which guarantees that for $n_{N_1}=n_{N_1}^{\rm eq}$ %and $N_{B-L}=0$
no asymmetry is
generated. Note that the old procedure for subtracting the RIS part of the $2\rightarrow 2$
process would have led to a contribution ${3\over 4} \g_{N_1}$ in the washout term rather
than ${1\over 2} \g_{N_1}$ \cite{gnx03}.
Neglecting the off-shell contribution $\g_{\D L=2}$, using the relation
\bea
\g_{N_1} =  n_{N_1}^{\rm eq}\ z\ H\ D = 2\ n_l^{\rm eq}\ z\ H\ W_{I\!D}\;,
\eea
and changing variables from $t$ to $z=M_1/T$ and from number densities to particle
numbers in a comoving volume (cf.~\cite{bdp02}), one obtains the Boltzmann equation
(\ref{dlg2}).

In order to obtain the kinetic equation (\ref{boltzm}) the correct
identification of the RIS term is essential. It is therefore of
crucial importance to derive this equation from first principles. In
the case of non-relativistic heavy neutrinos, i.e. $T < M_1$, this
has been done in \cite{bf00}. Note that in the strong washout regime,
where $T_B < M_1$, the decaying neutrinos are indeed non-relativistic.
Eq.~(49) of \cite{bf00} gives the
analogue of (\ref{boltzm}) for Boltzmann distribution functions rather
than the integrated number densities. The starting point of this
derivation are the Kadanoff-Baym equations which describe the full
quantum mechanical problem. Leptogenesis is then studied as a process
close to thermal equilibrium. As a consequence, the deviations of
distribution functions from equilibrium distribution functions, $\d
f_N(t,p)$ and $\d f_l(t,k)$ appear from the beginning\footnote{The
part of the Lagrangian involving left- and right-handed neutrinos has
a U(1) symmetry which implies $\d f_l(t,k) = - \d f_{\f}(t,k)$. Here
we have assumed that due to the other interactions in the standard
model $\d f_{\f}(t,k)=0$.}. For simplicity, in Eq.~(49) of ref.~\cite{bf00} the
contribution to the washout term from interferences with the heavy
neutrinos $N_2$ and $N_3$ has been neglected. Otherwise the result is
identical to Eq.~(\ref{boltzm}). In particular, the relative size of
the driving term for the asymmetry, which is proportional to $\ve_1$, and the
washout term due to inverse decays agrees with (\ref{boltzm}). It is important
to derive the Boltzmann equations and the reaction densities within a full
quantum mechanical treatment also for relativistic heavy neutrinos, in particular
in the resonance region $T \sim M_1$.

\begin{figure}[t]
\centerline{\psfig{file=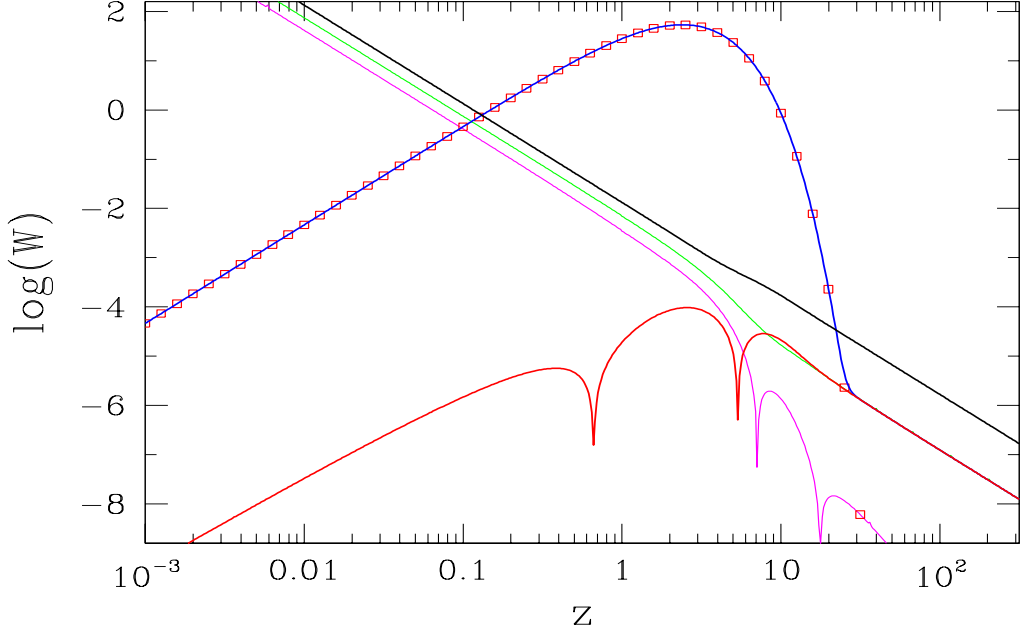,height=9cm,width=14cm}}
\caption{Absolute values of the different contributions to the washout term $W$
from the averaged squared matrix element in eq.~(\ref{DeltaLfinal}) for
$M_1=10^{10}\,$GeV, $\overline{m}=0.05\,$eV, $m_1=\overline{m}/\sqrt{3}$ and
$\widetilde{m}_1=0.03\,$eV. The solid
line is the term proportional to $1/D_1^2+1/R_1^2$, the black squares the contribution
from the delta function, the dotted line the term proportional to $\widetilde{m}_1^2$
in the second and third lines of eq.~(\ref{DeltaLfinal}), the dashed line the term
proportional to $m_1^2$, and the dashed-dotted line is the term proportional
to $\overline{m}^2$.\label{truth}}
\end{figure}

It is instructive to discuss the different contributions to $\gamma_{\Delta L=2}$,
the reaction density corresponding to the averaged matrix element squared
\begin{eqnarray}
 \left|{\cal M}_{\Delta L=2}\right|^2&=&
   2\,\left|{\cal M}_{\Delta L=2}\right|^2_+
   +2\,\left|{\cal M}_{\Delta L=2,t}\right|^2\nonumber\\[1ex]
&=& {4s^2\over v^4}\left\{\mt^2\left({1\over D_1^2}+{1\over R_1^2}\right)
    -32\pi^2{\mt v^2\over M_1}{1\over x^2}\delta(x-1)
    +24 \mb^2 \right.
    \label{DeltaLfinal}\\[1ex]
&& +\;\left.\mt^2\left[6+{8\over x}+{2\over x+1}
     +{1\over D_1}\left(3+{2\over x}\right)\right.\right.
    \nonumber\\[1ex]
&& \left.-\;{2\over x^2}\left(6x+3+{2\over x+2}+{x+1\over D_1}\right)
   {\rm ln}(x+1)\right]\nonumber\\[1ex]
&&  \left.-\;6\,m_1^2\left[2+{1\over x}
    +{1\over 2D_1}-{(2x+1)\over x^2}
    \ln(x+1)\right]\right\}\;,\nonumber
\end{eqnarray}
where we have again assumed the relation (\ref{int}). The different contributions
to the washout term $W$ are shown in Fig.~\ref{truth}. The term proportional
to $\overline{m}^2$, as well as the contributions from the last three lines
at high temperatures, $M_1\ll T\ll M_{2,3}$, give a simple power law behavior,
corresponding to Eqs.~(\ref{lowenergy}) and (\ref{intermediate}). At low temperatures,
the term proportional to $m_1^2$ rapidly approaches zero and becomes negligible.

It can be seen clearly
that for $z\lesssim30$ the contributions from the first two terms in Eq.~(\ref{DeltaLfinal})
cancel each other to a very good approximation, corresponding to the subtraction
of RIS contributions. However, the term proportional to $1/D_1^2+1/R_1^2$ has a
different low temperature limit than the delta function and cancels against the
term in the second and third lines of Eq.~(\ref{DeltaLfinal}) for $z\gtrsim30$.

Finally, this discussion is only applicable if off-shell and RIS contributions
can be separated. This is related
to the usual approximation that the right handed neutrinos can be considered
as asymptotic free states, i.e.\ that one can write down a Boltzmann equation
for them, which is the case if their width
is small, i.e.,
\begin{equation}
  c={\Gamma_1\over M_1}<\delta\;,
\end{equation}
where $\delta$ is some constant smaller than one. This translates
into the following condition for $\widetilde{m}_1$:
\begin{equation}
  \widetilde{m}_1 < \delta\, 0.76\,{\rm eV}
  \left({10^{15}\,{\rm GeV}\over M_1}\right)\;.
  \label{bound}
\end{equation}
We have checked numerically that the separation of on-shell and off-shell
contributions works well, as long as Eq.~(\ref{bound}) with $\delta=0.1$
is fulfilled.

\section*{Appendix B}

The scattering rates
are expressed through the reaction densities rates and these, in turn, through
the reduced cross sections,
\bea
\Gamma_{\phi,t(s)}^{(N_1)}=
{\gamma_{\phi,t(s)}\over n_{N_1}^{\rm eq}}={M_1\over 32\,g_{N_1}\,\pi^2}
\,{{\cal I}_{\phi,t(s)}(z)\over K_2(z)\,z^3} \,\, ,
\eea
where we introduced the following integrals
\begin{equation}
{\cal I}_{\phi,t(s)}(z)=\int_{z^2}^{\infty}\,d\psi\,\hat{\sigma}_{\phi,t(s)}(\psi)\,
\sqrt{\psi}\,K_1(\sqrt{\psi}) \,\, .
\end{equation}
The reduced cross sections can be written in the following form \cite{plu98}:
\begin{equation}
\hat{\sigma}_{\phi,t(s)}={3\,\alpha_{\mu}\over 4\,\pi}\,{M_1\,\tilde{m}_1\over v^2}\,
f_{\phi,t(s)}(x)
\end{equation}
with $ x\equiv{\psi/ z^2}$, $\alpha_{\mu}=m^2_t/v^2$
and where we defined the following functions:
\begin{equation}
f_{\phi,t}(x)=
{x-1\over x}\left[
        {x-2+2a_h\over x-1+a_h}+{1-2a_h\over x-1}
        \ln\left({x-1+a_h\over a_h}\right)\right]  \,\, ,
\end{equation}
\begin{equation}
f_{\phi,s}(x)=\left({x-1\over x}\right)^2 \,\, .
\end{equation}
with $a_h=(M_{\rm h}/M_1)^2$. The functions $f_{\phi,t(s)}(z)$
are then defined as:
\begin{equation}
f_{\phi,t(s)}(z)={\int_{z^2}^{\infty}\,d\psi\,f_{\phi,t(s)}({\psi/ z^2})\,
\sqrt{\psi}\,K_1(\sqrt{\psi})\over z^2\,K_2(z)} \,\, ,
\end{equation}
and in this way the Eq. (\ref{St}) for $S_t$ follows.

Similarly to the Eq. (\ref{approx}) for $K_2(z)$,
the modified Bessel function $K_3(z)$ can be approximated by
the analytical expression
\bea
K_3(z) \simeq {1\over z^3}\,\sqrt{1+{\pi\over 2}\,z}\,e^{-z}\,
\left({945\over 128}+{35\over 8}\,z+z^2 \right).
\eea
For $M_h/M_1 = 10^{-5}$ and small $K$, $z_{\rm eq}$ is well described by
\bea
z_{\rm eq} = 0.4 + 1.3\,\ln(1+K^{-0.88})\;.
\eea


\newpage

\begin{thebibliography}{99}

\bibitem{fy86}
M.~Fukugita, T.~Yanagida, \pl{174}{1986}{45}

\bibitem{bdp03}
W.~Buchm\"uller, P.~Di~Bari, M.~Pl\"umacher, \np{665}{2003}{445}

\bibitem{kt}
E.~W.~Kolb, M.~S.~Turner, {\it The Early Universe}, Addison-Wesley,
New York, 1990

\bibitem{kw80}
E.~W.~Kolb, S.~Wolfram, \np{172}{1980}{224}; \np{195}{1982}{542}(E)

%\bibitem{fot80}
%J.~N.~Fry, K.~A.~Olive, M.~S.~Turner, \pr{22}{1980}{2953}

\bibitem{hkx82}
J.~A.~Harvey, E.~W.~Kolb, D.~B.~Reiss, S.~Wolfram, \np{201}{1982}{16}

\bibitem{lut92}
M.~A.~Luty, \pr{45}{1992}{455}

\bibitem{plu97}
M.~Pl\"umacher, Z.~Phys.~{\bf C\ 74} (1997) 549

\bibitem{plu98}
M.~Pl\"umacher, \np{530}{1998}{207}

\bibitem{bcx00}
R.~Barbieri, P.~Creminelli, A.~Strumia, N.~Tetradis, \np{575}{2000}{61}

\bibitem{bdp02}
W.~Buchm\"uller, P.~Di~Bari, M.~Pl\"umacher, \np{643}{2002}{367}

\bibitem{pu03}
A.~Pilaftsis and T.~E.~J.~Underwood, hep-ph/0309342

\bibitem{gnx03}
G.~F.~Giudice, A.~Notari, M.~Raidal, A.~Riotto, A.~Strumia, hep-ph/0310123

\bibitem{ks88}
S.~Yu.~Khlebnikov, M.~E.~Shaposhnikov, \np{308}{1988}{885};\\
J.~A.~Harvey, M.~S.~Turner, \pr{42}{1990}{3344}

\bibitem{crx98}
L.~Covi, N.~Rius, E.~Roulet, F.~Vissani, \pr{57}{1998}{93}

\bibitem{ft81}
J.~N.~Fry, M.~S.~Turner, \pr{24}{1981}{3341}

\bibitem{lambert}
R.~M.~Corless et al., Adv.\ Comp.\ Math., Vol.~5 (1996) 329

\bibitem{lambert2}
F.~Chapeau-Blondeau and A.~Monir,
  IEEE Trans.\ Signal Processing, Vol.~50 (2002), 2160

\bibitem{news}
P.~Di Bari, AIP Conf.\ Proc.\  {\bf 655} (2003) 208
[hep-ph/0211175]
%%CITATION = HEP-PH 0211175;%%

\bibitem{fhy02}
M.~Fujii, K.~Hamaguchi, T.~Yanagida, \pr{65}{2002}{115012}

\bibitem{WMAP}
WMAP Collaboration, D.~N.~Spergel {\it et al.},
Astrophys.\ J.\ Suppl.\  {\bf 148} (2003) 175.

\bibitem{SDSS}
Max Tegmark {\it et al.}, astro-ph/0310723.

\bibitem{hmy02}
K.~Hamaguchi, H.~Murayama, T.~Yanagida, \pr{65}{2002}{043512}

\bibitem{phase03}
For recent discussions and references, see\\
Z.~Z.~Xing, hep-ph/0307359;\\
G.~C.~Branco, hep-ph/0309215;\\
W.~Rodejohann, hep-ph/0311142;\\
S.~Davidson, R.~Kitano, hep-ph/0312007

\bibitem{di02}
S.~Davidson, A.~Ibarra, \pl{535}{2002}{25}

\bibitem{bari}
G.~L.~Fogli, E.~Lisi, A.~Marrone, D.~Montanino, A.~Palazzo, A.~M.~Rotunno,
%``Neutrino oscillations: A global analysis,''
hep-ph/0310012.

\bibitem{atm}
%%CITATION = HEP-PH 0310012;%%
M.~H.~Ahn et al., K2K Collaboration,
%``Indications of neutrino oscillation in a 250-km long-baseline  experiment,''
\prl{90}{2003}{041801};\\
%[hep-ex/0212007];
%%CITATION = HEP-EX 0212007;%%
M. Shiozawa et al., SK Collaboration in {\em Neutrino 2002}, Proc. to appear.

\bibitem{solar}
Q.R. Ahmad {\it et al}, SNO Collaboration, nucl-ex/0309004;\\
K.~Eguchi {\it et al.}, KamLAND Collaboration,
%``First results from KamLAND: Evidence for reactor anti-neutrino  disappearance,''
\prl{90}{2003}{021802}

%%CITATION = HEP-EX 0212021;%%

\bibitem{ct03}
P.~H.~Chankowski, K.~Turzy\'nski, \pl{570}{2003}{198}

\bibitem{fps95}
M.~Flanz, E.~A.~Paschos, U.~Sarkar, \pl{345}{1995}{248};
\pl{384}{1996}{487} (E)

\bibitem{crv96}
L.~Covi, E.~Roulet, F.~Vissani, \pl{384}{1996}{169}

\bibitem{gkr01}
G.~F.~Giudice, E.~W.~Kolb and A.~Riotto, \pr{64}{2001}{023508}

\bibitem{hlx03}
T.~Hambye, Y.~Lin, A.~Notari, M.~Papucci, A.~Strumia, hep-ph/0312203

\bibitem{bdp022}
W.~Buchm\"uller, P.~Di~Bari, M.~Pl\"umacher, \pl{547}{2002}{128}

\bibitem{akx03}
S.~Antusch, J.~Kersten, M.~Lindner, M.~Ratz, \np{674}{2003}{401}

\bibitem{bp01}
W.~Buchm\"uller, M.~Pl\"umacher, \pl{511}{2001}{74}

\bibitem{bp00}
W.~Buchm\"uller, M.~Pl\"umacher, Int.~J.~Mod.~Phys. {\bf A 15} (2000) 5047

\bibitem{ls91}
G.~Lazarides, Q.~Shafi, \pl{258}{1991}{305};\\
H.~Murayama, T.~Yanagida, \pl{322}{1994}{349}

\bibitem{bf00}
W.~Buchm\"uller, S.~Fredenhagen, \pl{483}{2000}{217}

\bibitem{bp98}
W.~Buchm\"uller, M.~Pl\"umacher, \pl{431}{1998}{354}

\bibitem{rcv98}
E.~Roulet, L.~Covi, F.~Vissani, \pl{424}{1998}{101}

\bibitem{ci01}
J.~A.~Casas, A.~Ibarra, \np{618}{2001}{171}

\bibitem{saw03}
R.~F.~Sawyer, hep-ph/0312158

\end{thebibliography}
\end{document}